\documentclass[a4paper,11pt]{article}
\pdfoutput=1 % if your are submitting a pdflatex (i.e. if you have % images in pdf, png or jpg format)

\usepackage{jheppub} % for details on the use of the package, please
\usepackage{latexsym}
\usepackage{amsthm}
\usepackage{amsmath}
\usepackage{graphicx}
\usepackage{subcaption}
\usepackage{mathtools}
\usepackage{slashed}
\usepackage{amssymb}
\usepackage{color}
\usepackage{braket}
\usepackage{scalerel,stackengine}
\newcommand*\diff{\mathop{}\!\mathrm{d}}

\newcommand{\nn}{\nonumber}

\newcommand{\be}{\begin{eqnarray}}
\newcommand{\ee}{\end{eqnarray}}
\newcommand{\ma}{\mathrm}
\newcommand{\ml}{\mathcal}
\newcommand{\bs}{\boldsymbol}
\newcommand{\Tr}{\mathrm{Tr}}

\DeclareMathOperator{\sign}{sign}
\newcommand\pig[1]{\scalerel*[5pt]{\big#1}{%
    \ensurestackMath{\addstackgap[1.5pt]{\big#1}}}}

\begin{document}
\title{Quarkonium Semiclassical Transport in Quark-Gluon Plasma: Factorization and Quantum Correction}

\author[1]{Xiaojun Yao}
\affiliation[1]{Center for Theoretical Physics, Massachusetts Institute of Technology, Cambridge, MA 02139, USA}

\author[2]{and Thomas Mehen}
\affiliation[2]{Department of Physics, Duke University, Durham, NC 27708, USA}

\emailAdd{xjyao@mit.edu}
\emailAdd{mehen@phy.duke.edu}
\preprint{MIT-CTP/5233}
\abstract{We study quarkonium transport in the quark-gluon plasma by using the potential nonrelativistic QCD (pNRQCD) effective field theory and the framework of open quantum systems. We argue that the coupling between quarkonium and the thermal bath is weak using separation of scales, so the initial density matrix of the total system factorizes and the time evolution of the subsystem is Markovian. We derive the semiclassical Boltzmann equation for quarkonium by applying a Wigner transform to the Lindblad equation and carrying out a semiclassical expansion. We resum relevant interactions to all orders in the coupling constant at leading power of the nonrelativistic and multipole expansions. The derivation is valid for both weakly coupled and strongly coupled quark-gluon plasmas. We find reaction rates in the transport equation factorize into a quarkonium dipole transition function and a chromoelectric gluon distribution function. For the differential reaction rate, the definition of the momentum dependent chromoelectric gluon distribution function involves staple-shaped Wilson lines. For the inclusive reaction rate, the Wilson lines collapse into a straight line along the real time axis and the distribution becomes momentum independent. The relation between the two Wilson lines is analogous to the relation between the Wilson lines appearing in the gluon parton distribution function (PDF) and the gluon transverse momentum dependent parton distribution function (TMDPDF). The centrality dependence of the quarkonium nuclear modification factor measured by experiments probes the momentum independent distribution while the transverse momentum dependence and measurements of the azimuthal angular anisotropy may be able to probe the momentum dependent one. We discuss one way to indirectly constrain the quarkonium in-medium real potential by using the factorization formula and lattice calculations. The leading quantum correction to the semiclassical transport equation of quarkonium is also worked out. The study can be easily generalized to quarkonium transport in cold nuclear matter, which is relevant for quarkonium production in eA collisions in the future Electron-Ion Collider.
}
 
\maketitle
\flushbottom

\section{Introduction}
\label{sect:intro}
Heavy quarkonium has been used as a probe of the quark-gluon plasma (QGP) in heavy ion collisions for many years. The basic idea is the static screening effect in the hot plasma \cite{Matsui:1986dk,Karsch:1987pv}: the real part of the attractive potential between the heavy quark-antiquark pair ($Q\bar{Q}$) is significantly suppressed at high temperature and quarkonium ``melts". Therefore, suppression of quarkonium production in heavy ion collisions can be used as a signature of the QGP formation. The melting temperature can be defined for each quarkonium state as the minimum temperature when the state becomes unbound in the plasma. Since all quarkonium states have varying sizes, they are influenced by the static screening differently and thus have distinct melting temperatures, ordered by their sizes (or binding energies). A sequential melting pattern is expected where shallower quarkonium states melt at lower temperatures.

However, the simple static screening picture is complicated by several other factors: cold nuclear matter effects, as well as  quarkonium dissociation and recombination inside the plasma. One cold nuclear matter effect is the nuclear modification of parton distribution functions (PDF) inside heavy nuclei. Dissociation and recombination are hot medium effects. It has been shown that static screening and dissociation can be generated simultaneously from thermal loop correction to the quarkonium in-medium propagator \cite{Laine:2006ns,Beraudo:2007ky}. Thus if the static screening is included in the study, dissociation should also be taken into account for consistency. Furthermore, if a quarkonium state can still exist as a well-defined bound state above $T_c\sim 150$ MeV, which is a rough estimate of the transition temperature from the QGP phase to the hadronic phase\footnote{The transition is smooth at zero baryon chemical potential. So the $150$ MeV is just a rough estimate. In reality, the transition may happen in a range of tens of MeV in temperature.}, then (re)generation of this quarkonium state inside the QGP should also be possible \cite{Thews:2000rj}. It is expected that deeply bound states can start to (re)generate at high temperatures and do not have to wait until $T_c$, when light particles hadronize.

Quarkonium suppression has been intensively investigated in experiments at both the Relativistic Heavy Ion Collider (RHIC) and Large Hadron Collider (LHC). Semiclassical transport equations that account for static screening, dissociation and recombination~\cite{Grandchamp:2003uw,Grandchamp:2005yw,Yan:2006ve,Liu:2009nb,Song:2011xi,Song:2011nu,Sharma:2012dy,Nendzig:2014qka,Krouppa:2015yoa,Chen:2017duy,Zhao:2017yan,Du:2017qkv,Aronson:2017ymv,Ferreiro:2018wbd,Yao:2018zrg}, as well as statistical hadronization models~\cite{Andronic:2007bi,Andronic:2007zu}, have been used to study quarkonium production in heavy ion collisions and achieved great success in phenomenology. Anisotropic effects in the QGP have also been explored~\cite{Dumitru:2007hy,Dumitru:2009fy,Bhaduri:2020lur}. Recently, we derived the semiclassical transport equations using nonrelativistic QCD in the potential regime and the open quantum system formalism. The validity of this approach relies on a hierarchy of scales: $M \gg Mv \gg Mv^2,\, T$, where $M$ is the heavy quark mass, $v$ is the velocity of the quarks in the quarkonium and $T$ is the temperature of the plasma~\cite{Yao:2018nmy}. Long after quarkonium suppression was proposed as a diagnostic for the existence of the QGP, several other observables such as jet quenching and elliptic flow have been studied which more convincingly establish the existence of the QGP (see Refs.~\cite{Wiedemann:2009sh,Qin:2015srf,Heinz:2013th,Gale:2013da} for recent reviews on each topic). We no longer need to use quarkonium as a probe to answer the question whether the QGP is formed in heavy ion collisions. Then the natural question to ask is what we can learn about the QGP from quarkonium measurements. Since now the hot medium effects contain static screening, dissociation and recombination, a simple answer seems implausible.

  In deep inelastic scattering (DIS), electrons are shot onto protons to probe the inner structure of proton. The reason why this works is the possibility of constructing factorization theorems in certain kinematic limits (for a general discussion of factorization, see Ref.~\cite{Collins:1989gx}). These factorization theorems express physical observables as convolutions of perturbatively calculable cross sections with parton distribution functions (PDF) that can be expressed as matrix elements of operators within the proton state. Thus measurements of these cross sections determine specific correlation functions within the proton.  
  Furthermore, different observables can probe different kinds of parton distributions of the proton. Studying these distributions such as the transverse momentum dependent (TMD) PDF will be among the central scientific goals of the future Electron-Ion Collider.

This then leads us to ask: what correlation functions are measured when we study the production of quarkonium within the QGP? The same question could be asked of quarkonium production within, say, cold nuclear matter. We will use effective field theory (EFT) and the open quantum system framework to answer this question. For processes involving light-like partons, Soft-Collinear Effective Theory has been widely applied to study factorization in various processes \cite{Bauer:2000yr,Bauer:2001yt,Bauer:2002nz,Manohar:2006nz,Fleming:2007xt,Chien:2015cka,Rothstein:2016bsq,Frye:2016aiz,Makris:2017arq,Gutierrez-Reyes:2019msa,Chien:2019osu}. In our case since the quarkonium is nonrelativistic we will use potential nonrelativistic QCD (potential NRQCD or pNRQCD) \cite{Brambilla:1999xf,Brambilla:2004jw,Fleming:2005pd}. This EFT has been used to study static screening and dissociation \cite{Brambilla:2008cx,Brambilla:2011sg,Brambilla:2013dpa}. Also, it has been used in a Lindblad equation to define a transport coefficient of quarkonium \cite{Brambilla:2019tpt}. New developments of nonrelativistic EFT for quarkonium can be found in Refs.~\cite{Makris:2019ttx,Fleming:2019pzj}. The open quantum system framework describes the dynamics of a subsystem, interacting with an environment. When the environment is traced over, the subsystem evolves as an open system. The open quantum system framework has been recently applied to study quarkonium transport in the QGP \cite{Yao:2018nmy,Young:2010jq,Borghini:2011ms,Akamatsu:2011se,Akamatsu:2014qsa,Blaizot:2015hya,Katz:2015qja,Brambilla:2016wgg,Brambilla:2017zei,Kajimoto:2017rel,DeBoni:2017ocl,Blaizot:2017ypk,Blaizot:2018oev,Akamatsu:2018xim,Miura:2019ssi} and provides new insight to our understanding of quarkonium transport. The open quantum system approach can be thought of as an extension of the Schr\"odinger equation with a complex potential~\cite{Petreczky:2010tk,Islam:2020gdv} and it takes into account both correlated and uncorrelated recombination consistently. The difference between correlated and uncorrelated recombination is discussed in Ref.~\cite{Yao:2020xzw}.

We will derive the semiclassical Boltzmann equation for quarkonium in the QGP by applying a Wigner transform (a Gaussian smearing is required for sustaining positivity) to the Lindblad equation, which is the evolution equation for the open system. The interaction between quarkonium (subsystem) and the thermal QGP (environment) is weak according to the power counting of pNRQCD under the assumed hierarchy. Therefore the density matrix of the total system can be assumed to factorize into the density matrix of the subsystem and that of the environment. Furthermore, in the weak coupling (between the subsystem and the environment) limit, the time evolution of the subsystem can be shown to be Markovian. We will work to leading order in the power counting parameter of pNRQCD but resum interactions to all orders in the strong coupling constant, which can be written in terms of Wilson lines. We will show that the reaction rates in the Boltzmann equation factorize into a quarkonium dipole transition function and a chromoelectric gluon distribution function of the thermal QGP. For the differential reaction rate, the chromoelectric gluon distribution function is momentum dependent and has spatially separated chromoelectric fields connected via a staple-shaped Wilson line extending to the time infinity. For the inclusive reaction rate, the function is momentum independent and the Wilson line collapses into a straight timelike line of finite length. The structures of the Wilson lines are very similar to the case of gluon TMDPDF and gluon PDF, respectively, except that here the Wilson lines lie along a timelike direction rather than the lightcone. The momentum independent chromoelectric gluon distribution function has been studied in Ref.~\cite{Brambilla:2019tpt} as one quarkonium transport coefficient and in Refs.~\cite{CasalderreySolana:2006rq,CaronHuot:2008uh} as the heavy quark diffusion coefficient. A recent lattice calculation of the heavy quark diffusion coefficient can be found in Ref.~\cite{Brambilla:2020siz}. What is new in this paper is the definition and discussion of the momentum dependent chromoelectric structure function. We will discuss one application of the factorized rate to constrain the in-medium real potential between a $Q\bar{Q}$ pair. A point of emphasis is that the factorization of the reaction rate crucially depends on the factorization of the density matrix into the subsystem density matrix and the environment density matrix. This is generally believed to be true in the weak coupling limit, where the weak coupling is between the subsystem and the environment. The subsystem and the environment themselves can be strongly coupled.

This paper extends our earlier work~\cite{Yao:2018nmy} which derived semiclassical transport equations from pNRQCD and the open quantum system formalism. Major improvements compared to Ref.~\cite{Yao:2018nmy} include:
\begin{itemize}
\item{We resum the $A_0$ interaction and the interaction mediated by Coulomb modes between the octet $Q\bar{Q}$ state and the thermal QGP. This allows us to define the chromoelectric gluon distribution function nonperturbatively and construct the factorization. The derivation of Ref.~\cite{Yao:2018nmy} only works for a weakly coupled QGP while the derivation we show here is also valid for a strongly coupled QGP. A formalism compatible with a strongly coupled QGP is crucial for phenomenological studies of the future experiments at RHIC and LHC, especially the one carried out by the sPHENIX collaboration since the plasma temperature achieved at RHIC is not very high.}

\item{We carry out a systematic semiclassical expansion for the recombination term. For dissociation, no semiclassical expansion is needed. The semiclassical expansion is a gradient expansion. We work out the leading quantum correction to the semiclassical Boltzmann equation.}
\end{itemize}

The paper is organized as follows: In Section~\ref{sect:pnrqcd}, we will discuss the hierarchy of scales in the problem and briefly explain pNRQCD with a focus on the power counting. Then in Section~\ref{sect:open}, a short introduction to the open quantum system framework will be given. Derivation of the semiclassical Boltzmann equation will be elucidated in detail in Section~\ref{sect:transport}. Factorization of the reaction rates will also be discussed there. We will derive the leading quantum correction to the Boltzmann equation in Section~\ref{sect:quantum}. Finally, we will summarize and draw conclusions in Section~\ref{sect:conclusion}.

\section{Separation of Scales and Potential NRQCD}
\label{sect:pnrqcd}
We consider the following hierarchy of scales: $M \gg Mv \gg Mv^2,\,T,\,\Lambda_{\text{QCD}}$, where $M$ is the heavy quark mass, $v$ is the typical relative velocity between the heavy quark pair inside quarkonium, $T$ is the temperature of the medium and $\Lambda_{\text{QCD}}$ is the nonperturbative scale of QCD. In vacuum, $M \gg Mv \gg Mv^2$ is the standard separation of scales for nonrelativistic heavy quarks \cite{Bodwin:1994jh}. For both charmonium and bottomonium, $Mv^2\sim 500$ MeV. In current heavy ion collision experiments, $T\sim 500$ MeV is roughly the highest temperature achieved in the early stage of the medium expansion. Naively, we would expect $Mv^2 \gtrsim T$ to be approximately valid during the whole lifetime of QGP. But due to the static plasma screening, the binding energy decreases as the temperature increases, and $Mv^2\sim 500$ MeV is probably no longer true except for the bottomonium ground state at $T\sim 500$ MeV. So we do not specify the hierarchy between $Mv^2$ and $T$. However, we still believe $Mv \gg T$ is a relevant hierarchy for quarkonium transport. The reason is the following: when the temperature is high and the quarkonium size is large, this hierarchy $Mv \gg T$ may be violated and the interaction between the heavy quark pair is significantly suppressed. Loosely bound quarkonium states at such a high temperature, even if they exist in the medium, are no longer well-defined bound states, the dissociation rate is very large and formation is ineffective and can be neglected. The state of the heavy quark and antiquark will be more like two open heavy quarks and their dynamics is governed by the transport of open heavy flavors rather than the transport of quarkonium. Only when the QGP expands and the temperature drops, does the quarkonium formation become effective and transport become applicable. At the end of the QGP expansion when hadronization starts to occur, the temperature is about $T_c \sim 150$ MeV and every quarkonium state should regain their vacuum properties. Therefore, the hierarchy $M \gg Mv \gg Mv^2 ,\, T,\,\Lambda_{\text{QCD}}$ should be valid for ground and lower excited quarkonium states in the later stage of the QGP expansion, when the formation of these states becomes effective.

Under the separation of scales $M \gg Mv \gg Mv^2,\,T,\,\Lambda_{\text{QCD}}$, one can construct an effective field theory of QCD, pNRQCD, by systematic nonrelativistic expansion in $v$ and multipole expansion in $rT \sim \frac{T}{Mv}$, where $r$ is the typical size of a quarkonium state.\footnote{When $T\sim Mv^2$, the multipole expansion is equivalent to the nonrelativistic expansion. Then $v$ is only power counting parameter.} In our assumed hierarchy of scales, the effective field theory can be constructed perturbatively at the scale $Mv$. The Lagrangian can be written as
\be
\label{eq:lagr}
\ml{L}_\ma{pNRQCD} &=& \int \diff^3r\, \Tr\Big(  \ma{S}^{\dagger}(i\partial_0-H_s)\ma{S} +\ma{O}^{\dagger}( iD_0-H_o )\ma{O} + V_A( \ma{O}^{\dagger}\bs r \cdot g{\bs E} \ma{S} + \ma{h.c.})  \nn \\
&&+ \frac{V_B}{2}\ma{O}^{\dagger}\{ \bs r\cdot g\bs E, \ma{O}  \} +\cdots \Big) + \ml{L}_\ma{light\ quark} + \ml{L}_\ma{gluon} \, ,
\ee
where higher order terms in the expansion are neglected. We will work at leading power in the nonrelativistic and multipole expansions throughout the paper. Here ${\bs E} = {\bs E}^A T^A$ ($A$ is the adjoint color index) represents the chromoelectric field and $D_0\ma{O} = \partial_0\ma{O} -ig [A_0, \ma{O}]$. The gluon and light quark parts are just QCD with momenta $\lesssim Mv$. The matching coefficients are $V_A=V_B=1$ at lowest order in the coupling constant at the scale $Mv$. The composite fields for the quarkonia are the color singlet $\ma{S}(\bs R, \bs r, t)$ and color octet $\ma{O}(\bs R, \bs r, t)$ where $\bs R$ denotes the center-of-mass (c.m.) position and $\bs r$ the relative coordinate. The composite fields here are constructed by projecting onto the proper color space, a heavy quark field and a heavy antiquark field at the same time, connected by a spatial Wilson line. We will assume the medium is invariant under translation so the existence of the medium does not break the separation into the c.m.~and relative motions. $H_{s}$ and $H_{o}$ denote the color singlet and octet Hamiltonians of the relative motion. At leading order in the nonrelativistic expansion, $H_{s,o} = -\frac{\nabla_{\bs r}^2}{M} + V_{s,o}({\bs r})$, where $V_{s,o}$ are the singlet and octet potentials and are attractive and repulsive, respectively. So only a color singlet $Q\bar{Q}$ pair can be bound. 
Both potentials are spin independent. Thus the pNRQCD Lagrangian is invariant under heavy quark spin symmetry and we will ignore spin quantum numbers in this paper. In the $v$ expansion, the Fock state $|Q\bar{Q}g\rangle$ of quarkonium, in which the $Q\bar{Q}$ pair is a color octet state, is suppressed by at least $v^2$ in probability with respect to the leading Fock state $|Q\bar{Q}\rangle$ in which the $Q\bar{Q}$ pair is a color singlet state. Therefore, at leading order in the $v$ expansion, which is the order we are working, quarkonium can only be a color singlet $Q\bar{Q}$ pair. Furthermore, the QGP is a deconfined phase of QCD, where light quarks and gluons are liberated. Thus it is reasonable to assume no gluon can be bound with an octet $Q\bar{Q}$ pair. 

For our construction that is at leading power in $v$ and $rT$ but resummed to all orders in $\alpha_s(Mv^2, T)$, we do the following redefinitions:
\be
\ma{O}(\bs R, \bs r, t) &\to& W_{[(\bs R,t),(\bs R, t_L)]} \widetilde{\ma{O}}(\bs R, \bs r, t) (W_{[(\bs R,t),(\bs R, t_R)]})^\dagger \nn \\
\label{eqn:O_redef}
&=& W_{[(\bs R,t),(\bs R, t_L)]} \widetilde{\ma{O}}(\bs R, \bs r, t) W_{[(\bs R, t_R),(\bs R,t)]}\,,
\ee
where the Wilson line in the fundamental representation is defined by
\be
\label{eqn:wilson_line}
 W_{[(\bs R, t_f),(\bs R, t_i)]} = \ml{P}\exp\Big( ig\int^{t_f}_{t_i} \diff s A_0({\bs R}, s)\Big) \,,
\ee
in which the path is a straight line from $(\bs R,t_i)$ to $(\bs R,t_f)$. In Eq.~(\ref{eqn:O_redef}), $t_L$ and $t_R$ are choices of the end points. If we want $\widetilde{O}$ to be invariant under a gauge transformation (i.e., to behave like a singlet operator), we need to set $t_L=t_R\equiv t_0$ to guarantee that both the left and right sides of Eq.~(\ref{eqn:O_redef}) transform in the same way under a gauge transformation. Then we can write Eq.~(\ref{eqn:O_redef}) in a simpler form by using Wilson lines in the adjoint representation
\be
\label{eqn:O_redef2}
\ma{O}(\bs R, \bs r, t) \to \ml{W}_{[(\bs R,t),(\bs R, t_0)]}
\widetilde{\ma{O}}(\bs R, \bs r, t)\,,
\ee
where $\ml{W}_{[(\bs R,t),(\bs R, t_0)]}$ denotes a Wilson line in the adjoint representation that has a straight path from $(\bs R, t_0)$ to $(\bs R,t)$. So far $t_0$ is an arbitrary constant, but later we will follow arguments given in Ref.~\cite{Arnesen:2005nk} and show the results are independent of the choice of $t_0$. Now we can simplify the octet kinetic term
\be \nn
&&D_0\ma{O}(\bs R, \bs r, t) = \partial_0\ma{O}(\bs R, \bs r, t) -ig A_0(\bs R, t)  \ma{O}(\bs R, \bs r, t) + ig \ma{O}(\bs R, \bs r, t) A_0(\bs R, t) \\[4pt]
%&=& W_{[(\bs R,t),(\bs R, t_L)]} \partial_0 \widetilde{\ma{O}}(\bs R, \bs r, t) W_{[(\bs R, t_R),(\bs R,t)]} -ig A_0(\bs R, t)  \ma{O}(\bs R, \bs r, t) + ig \ma{O}(\bs R, \bs r, t) A_0(\bs R, t)\\[4pt]  \nn
%&& + ig A_0(\bs R,t) W_{[(\bs R,t),(\bs R, t_L)]} \widetilde{\ma{O}}(\bs R, \bs r, t) W_{[(\bs R, t_R),(\bs R,t)]} \\[4pt]\nn 
%&& -ig W_{[(\bs R,t),(\bs R, t_L)]} \widetilde{\ma{O}}(\bs R, \bs r, t) W_{[(\bs R, t_R),(\bs R,t)]} A_0(\bs R,t)  \\[4pt] 
&=& W_{[(\bs R,t),(\bs R, t_0)]} \partial_0 \widetilde{\ma{O}}(\bs R, \bs r, t) W_{[(\bs R, t_0),(\bs R,t)]} = \ml{W}_{[(\bs R,t),(\bs R, t_0)]}
\partial_0\widetilde{\ma{O}}(\bs R, \bs r, t)\,.
\ee
The octet kinetic term can then be rewritten as
\be
\int \diff^3r \Tr\Big( \widetilde{\ma{O}}^{\dagger} (i\partial_0 - H_o) \widetilde{\ma{O}}  \Big)\,.
\ee
The new dipole interaction between the singlet and octet is given by
\be
g \int \diff^3r \Tr\Big( \widetilde{\ma{O}}^{\dagger} r_i   \ml{W}_{[(\bs R, t_0),(\bs R,t)]}  E_i  \ma{S} + \ma{S}^\dagger r_i  E_i  \ml{W}_{[(\bs R, t),(\bs R,t_0)]} \widetilde{\ma{O}}  \Big) \,. \ \ \ \ 
\ee
If we define
\be
\label{eqn:E}
\widetilde{E}_i(\bs R,t) &\equiv& \ml{W}_{[(\bs R, t_0),(\bs R,t)]}  E_i(\bs R,t)  \\[4pt] 
\label{eqn:Edagger}
\widetilde{E}^{\dagger}_i(\bs R,t) &\equiv&  E_i(\bs R,t)  \ml{W}_{[(\bs R, t),(\bs R,t_0)]} \,,
\ee
the dipole interaction term can be written as
\be
g \int \diff^3r \Tr\Big( \widetilde{\ma{O}}^{\dagger} r_i   \widetilde{E}_i \ma{S} + \ma{S}^\dagger r_i  \widetilde{E}^{\dagger}_i \widetilde{\ma{O}}  \Big) \,.
\ee

For later convenience, we introduce the ``bra-ket" notation\footnote{The prefactors of $\frac{1}{\sqrt{N_c}}$ and $\frac{1}{\sqrt{T_F}}$ are introduced such that the kinetic terms in the Lagrangian have unit prefactors after the trace is written out explicitly}:
\be 
\langle {\bs r} | S({\bs R}, t) \rangle &\equiv& S({\bs R}, {\bs r}, t) \equiv \frac{1}{\sqrt{N_c}} \ma{S}({\bs R}, {\bs r}, t) \\ 
\langle {\bs r} | {O}^A({\bs R}, t) \rangle  &\equiv& {O}^A({\bs R}, {\bs r}, t) \equiv \frac{1}{\sqrt{T_F}}\Tr[T^A {\ma{O}}({\bs R}, {\bs r}, t) ] \\
\langle {\bs r} | \widetilde{O}^A({\bs R}, t) \rangle  &\equiv& \widetilde{O}^A({\bs R}, {\bs r}, t) \equiv \frac{1}{\sqrt{T_F}}\Tr[T^A \widetilde{\ma{O}}({\bs R}, {\bs r}, t) ] \\
\langle S({\bs R}, t) | f({\bs r}) | \widetilde{O}^A({\bs R}, t) \rangle &\equiv& \int\diff^3r S^\dagger({\bs R}, {\bs r}, t) f({\bs r}) \widetilde{O}^A({\bs R}, {\bs r}, t)\,,
\ee
for any function $f$ of ${\bs r}$. At the current leading power calculation, we will use $f({\bs r}) = r_i$ where $i=x,y,z$. Here $N_c=3$ and $T_F=\frac{1}{2}$, which is defined by $\Tr[T^A T^B] = T_F \delta^{AB}$. For later use, we define $C_F = \frac{N_c^2-1}{2N_c}$.
The quantization of the fields is given by
\be \nn
|S(\bs R, t) \rangle &=& \int\frac{\diff^3 p_\ma{cm}}{(2\pi)^3}  e^{-i(Et-\bs p_\ma{cm} \cdot \bs R)} \bigg( \sum_{nl} a_{nl}(\bs p_\ma{cm}) \otimes | \psi_{nl} \rangle   + \int\frac{\diff^3 p_{\ma{rel}}}{(2\pi)^3} b_{{\bs p}_{\ma{rel}}}(\bs p_\ma{cm}) \otimes | \psi_{{\bs p}_{\ma{rel}}} \rangle \bigg) \\ 
\label{eqn:O_quanti}
| \widetilde{O}^A(\bs R, t) \rangle &=&  \int\frac{\diff^3 p_\ma{cm}}{(2\pi)^3} e^{-i(Et-\bs p_\ma{cm}\cdot \bs R)}  \int\frac{\diff^3 p_{\ma{rel}}}{(2\pi)^3} c^A_{{\bs p}_{\ma{rel}}}(\bs p_\ma{cm}) 
\otimes | \Psi_{{\bs p}_{\ma{rel}}} \rangle \,,
\ee
where $E$ is the eigenenergy of a state that will be explained below. The operators $a^{(\dagger)}_{nl}(\bs p_\ma{cm})$, $b^{(\dagger)}_{{\bs p}_{\ma{rel}}}(\bs p_\ma{cm})$ and $c^{A(\dagger)}_{{\bs p}_{\ma{rel}}}(\bs p_\ma{cm})$ act on the Fock space to annihilate (create) composite particles with the c.m.~momentum ${\bs p_\ma{cm}}$ and the corresponding quantum numbers in the relative motion. These quantum numbers can be $nl$ for bound singlet states, ${\bs p}_{\ma{rel}}$ for unbound singlet states and color $A$ and ${\bs p}_{\ma{rel}}$ for unbound octet states. When we compute the matrix elements, we will average over the polarizations of non-$S$ wave quarkonium states. So in our notation, we omit the quantum number $m_l$ of the bound singlet state. In the octet channel no bound state exists because of the repulsive octet potential. The corresponding wavefunctions of the relative motion are $| \psi_{nl} \rangle$ and $| \psi_{{\bs p}_{\ma{rel}}} \rangle$ for color-singlet states and $ | \Psi_{{\bs p}_{\ma{rel}}} \rangle $ for color-octet states. They can be obtained by solving the equations of motion of the free composite fields, which are Schr\"odinger equations. The eigenenergies are $E = -|E_{nl}|$ and $E = \frac{{\bs p}_{\ma{rel}}^2}{M}$ for the bound and unbound states, respectively, with higher order terms in $v$ neglected. Here $E_{nl}$ is the binding energy of the bound state $| \psi_{nl} \rangle$.

The dissociation and recombination of quarkonium occur via the dipole interaction between the color singlet and octet states. As explained above, quarkonium is treated as a color singlet $Q\bar{Q}$ pair in this work, consistent with the leading power (in $v$) calculation. The dipole vertex scales as $rT$ where $r\sim\frac{1}{Mv}$ is the typical quarkonium size and $T$ is the typical energy and momentum scale of excitations in the plasma (the excitation energy and momentum comes from the derivative in the chromoelectric field). In our assumed hierarchy $Mv \gg  T$, the dipole vertex between the singlet and octet states scales as $rT \ll 1$. Thus, the interaction between quarkonium and the medium is weak. This argument is about the weak coupling between quarkonium and the QGP and is valid even if quarkonium and the QGP are strongly coupled themselves.
%The argument is based on a separation of scales and thus is valid even if the pNRQCD is constructed nonperturbatively. 
%Perturbatively at lowest order in $\alpha_s$,  $V_A=V_B=1$  in the Lagrangian (\ref{eq:lagr}) and we will set $V_A=V_B=1$ in this paper for simplicity. The results can be easily modified for nontrivial $V_A$ and $V_B$.

\section{Open Quantum Systems}
\label{sect:open}
In this section, we briefly introduce the framework of open quantum systems. We consider a total system consisting of a subsystem and an environment (thermal bath). The total Hamiltonian is given by
\be
H = H_S +H_E + H_I\,,
\ee
where $H_S$ is the subsystem Hamiltonian, $H_E$ is the environment Hamiltonian, and $H_I$ contains the interactions between the subsystem and the environment. The interaction Hamiltonian is assumed to be factorized as follows: $H_I = \sum_{\alpha} O^{(S)}_{\alpha} \otimes O^{(E)}_{\alpha}$ where $\alpha$ denotes all quantum numbers. (For local quantum field theory, this is generally true and $\alpha$ includes the spatial coordinates.) The operators $O^{(S)}_{\alpha}$ are of the subsystem while $O^{(E)}_{\alpha}$ are of the environment. We can assume $\langle O^{(E)}_{\alpha}\rangle \equiv \Tr_E(O^{(E)}_{\alpha}\rho_E) = 0$ because we can redefine $O^{(E)}_{\alpha}$ and $H_S$ by $O^{(E)}_{\alpha} - \langle O^{(E)}_{\alpha}\rangle$ and $H_S + \sum_{\alpha} O^{(S)}_{\alpha} \langle O^{(E)}_{\alpha}\rangle $ respectively. Here $\rho_E$ is the density matrix of the environment. Each part of the Hamiltonian is assumed to be Hermitian.

The von Neumann equation for the time evolution of the density matrix in the interaction picture is given by
\be
\frac{\diff \rho^{(\ma{int})}(t)}{\diff t} = -i [H^{(\ma{int})}_I(t), \rho^{(\ma{int})}(t)] \,.
\ee 
We will omit the superscript ``(int)" in the following. The symbolic solution is given by
\be
\rho(t) = U(t,0) \rho(0) U^{\dagger}(t,0)\,,
\ee
where the evolution operator is
\be
U(t,0) = \ml{T} e^{-i \int_0^t H_I(t') \diff t'}\,,
\ee
and $\ml{T}$ is the time-ordering operator. The starting time $t=0$ is arbitrary in the Markovian limit. So for later convenience, we will shift the starting time to $-t/2$ and obtain
\be
\rho\Big(\frac{t}{2}\Big) = U\Big(\frac{t}{2},\frac{-t}{2}\Big) \rho\Big(\frac{-t}{2}\Big) U^{\dagger}\Big(\frac{t}{2},\frac{-t}{2}\Big)\,.
\ee

We will assume the subsystem and the environment are weakly interacting. This is valid for quarkonium inside the QGP with the hierarchy of scales $M \gg Mv \gg Mv^2 ,\, T ,\, \Lambda_{\text{QCD}}$ because the heavy quark pair interacts with the medium chromoelectric field via the dipole interaction, which scales as $rT \sim \frac{T}{Mv} \ll 1$. Then we can assume the initial density matrix factorizes
\be
\rho\Big(\frac{-t}{2}\Big) = \rho_S\Big(\frac{-t}{2}\Big) \otimes \rho_E\,,
\ee
which is generally true for weakly coupled systems (factorization breaking terms come at higher orders in the coupling). The environment density matrix is assumed to be in thermal equilibrium thus  $\rho_E = \frac{1}{Z} e^{-\beta H_E}$ where $\beta = 1/T$, and is time-independent. We will use $\langle O^{(E)}\rangle_T$ to denote $\Tr_E[O^{(E)}\rho_E]$ from now on, where the subscript $T$ indicates the environment is a thermal bath at temperature $T$. If we expand the interaction to second order in perturbation (which corresponds to leading power in $rT$) and take the partial trace over the environment degrees of freedom, we obtain the Lindblad equation:
\be 
\rho_S\Big(\frac{t}{2}\Big) &=& \rho_S\Big(\frac{-t}{2}\Big)  -i\sum_{a,b} \sigma_{ab}(t) \Big[L_{ab}, \rho_S\Big(\frac{-t}{2}\Big) \Big]  \nn \\
\label{eqn:lindblad}
& + & \sum_{a,b,c,d} \gamma_{ab,cd} (t) \Big( L_{ab} \rho_S\Big(\frac{-t}{2}\Big) L^{\dagger}_{cd} - \frac{1}{2} \Big\{ L^{\dagger}_{cd}L_{ab}, \rho_S\Big(\frac{-t}{2}\Big)\Big\}  \Big)
 + \ml{O}(H_I^3)\,.
\ee
Each term is defined as
\be
L_{ab} &\equiv& |a\rangle \langle b| \\[4pt] 
\sigma_{ab}(t) &\equiv& \frac{-i}{2}  \sum_{\alpha, \beta} \int_{\frac{-t}{2}}^{\frac{t}{2}} \diff t_1 \int_{\frac{-t}{2}}^{\frac{t}{2}} \diff t_2 \Tr_E \Big[ \sign(t_1-t_2) \langle a | O^{(S)}_{\alpha}(t_1) O^{(S)}_{\beta}(t_2) | b\rangle C_{\alpha\beta}(t_1, t_2) \Big] \ \ \ \ \ \ \\
\gamma_{ab,cd} (t) &\equiv&  \sum_{\alpha, \beta} \int_{\frac{-t}{2}}^{\frac{t}{2}} \diff t_1 \int_{\frac{-t}{2}}^{\frac{t}{2}} \diff t_2 \Tr_E\Big[  \langle d | O^{(S)}_{\alpha}(t_1) | c \rangle \langle a | O^{(S)}_{\beta}(t_2) | b\rangle C_{\alpha\beta}(t_1, t_2) \Big] \\[4pt]
C_{\alpha\beta}(t_1, t_2) &\equiv& O^{(E)}_{\alpha}(t_1)O^{(E)}_{\beta}(t_2)\rho_E
\,,
\ee
where $\{ |a\rangle \}$ forms a complete set of states in the Hilbert space of the subsystem. We can choose them to be the set of eigenstates of $H_S$.

To calculate these in pNRQCD at finite temperature, we need the following dictionary:
\be
O^{(S)}_{\alpha} &\to& \langle S(\bs R, t) | r_i | \widetilde{O}^A(\bs R, t)\rangle\,,\ \ \langle \widetilde{O}^{A}(\bs R, t) | r_i | S(\bs R, t)\rangle \\[4pt] 
O^{(E)}_{\alpha} &\to& g\sqrt{\frac{T_F}{N_c}} \widetilde{E}_i^{A\dagger}(\bs R, t)\,, \ \ g\sqrt{\frac{T_F}{N_c}} \widetilde{E}_i^{A}(\bs R, t) \\[4pt] 
\sum_{\alpha} &\to& \int \diff^3 R \sum_i \sum_A\,,
\ee
where $i=x,y,z$ is the spatial component and the superscript $A$ denotes the adjoint color index. The complete set of states $|a\rangle$ in pNRQCD is
\be 
 \label{eqn:s_create}
|\bs k, nl, 1\rangle &=& a^{\dagger}_{nl}(\bs k) | 0 \rangle \\[4pt] 
 |{\bs p}_{\ma{cm}}, {\bs p}_{\ma{rel}} ,1\rangle  &=& b^{\dagger}_{{\bs p}_{\ma{rel}}}({\bs p}_{\ma{cm}}) | 0 \rangle\\[4pt] 
 \label{eqn:octet_create}
 |{\bs p}_{\ma{cm}}, {\bs p}_{\ma{rel}}, A\rangle(t)  &=& c^{B\dagger}_{{\bs p}_{\ma{rel}}}({\bs p}_{\ma{cm}}) \ml{W}^{BA}_{[t_0,t]} | 0 \rangle\,,
\ee
where the label $1$ means the state is a color singlet while $A$ means the state is in a specific color octet state. Since the redefined octet field is dressed with a Wilson line, Eq.~(\ref{eqn:octet_create}) contains a Wilson line for the octet state created by the original octet field $ |{\bs p}_{\ma{cm}}, {\bs p}_{\ma{rel}}, A\rangle(t)$, where $c^{B\dagger}_{{\bs p}_{\ma{rel}}}({\bs p}_{\ma{cm}})$ is creation operator for the redefined octet field.

With these preparations, we are ready to derive the semiclassical transport equation and construct factorized reaction rates.

\section{Transport Equation and Factorized Rates}
\label{sect:transport}
We will first sketch the derivation of the semiclassical transport equation and then explain it in detail. 
The semiclassical transport equation for quarkonium can be obtained from the Lindblad equation by first making the Markovian approximation, which is valid when the environment correlation time is much smaller than the subsystem relaxation time. The Markovian approximation is generally true when the subsystem interacts weakly with the environment. The environment correlation time is given by $\sim 1/T$ while the subsystem relaxation time can be estimated as $\sim \frac{1}{(rT)^2T} \sim \frac{(Mv)^2}{T^3}$, where we use the dipole interaction between quarkonium and the medium as the relaxation process. It can be seen that the Markovian approximation is valid in our power counting. Under the Markovian approximation, the Lindblad equation (\ref{eqn:lindblad}) in the Schr\"odinger picture when $t$ is small turns to \cite{Yao:2018nmy}
\be
\label{eqn:preBoltzmann}
f_{nl}\Big({\bs x}, {\bs k}, \frac{t}{2}\Big) &=& f_{nl}\Big({\bs x}, {\bs k}, \frac{-t}{2}\Big) \nn\\
&-& t \frac{{\bs k}}{2M} \cdot \nabla_{\bs x}  f_{nl}\Big({\bs x}, {\bs k}, \frac{-t}{2}\Big) +  t\ml{C}_{nl}^+\Big({\bs x}, {\bs k}, \frac{-t}{2}\Big) - t\ml{C}_{nl}^-\Big({\bs x}, {\bs k}, \frac{-t}{2}\Big) \,,
\ee
if a Wigner transform is applied to the subsystem density matrix\footnote{The phase space distribution defined in this way is not generally positive-definite, but one can make it positive-definite by a Gaussian smearing.}
\be
\label{eqn:wigner}
f_{nl}({\bs x}, {\bs k}, t) \equiv \int\frac{\diff^3k'}{(2\pi)^3} e^{i {\bs k}'\cdot {\bs x} } \Big\langle  {\bs k}+\frac{{\bs k}'}{2}, nl,1   \Big| \rho_S(t)  \Big|   {\bs k}-\frac{{\bs k}'}{2} , nl, 1 \Big\rangle \,.
\ee
Dividing (\ref{eqn:preBoltzmann}) by $t$ and taking the limit $t\to0$, we obtain the Boltzmann transport equation at $t=0$. Since the starting time is arbitrary, we interpret a similar equation at an arbitrary time $t$ as
\be
\label{eqn:Boltzmann}
\frac{\partial}{\partial t} f_{nl}({\bs x}, {\bs k}, t) + \frac{{\bs k}}{2M} \cdot \nabla_{\bs x} f_{nl}({\bs x}, {\bs k}, t)
= \ml{C}_{nl}^+({\bs x}, {\bs k}, t) - \ml{C}_{nl}^-({\bs x}, {\bs k}, t) \,.
\ee
The Boltzmann equation describes the phase-space evolution of the quarkonium state with the quantum number $nl$. To solve this equation, one needs to couple it with transport equations for unbound singlet and octet $Q\bar{Q}$'s, since the recombination term depends on the distribution of unbound $Q\bar{Q}$'s (see Section~\ref{sect:semi}).

In Eqs.~(\ref{eqn:preBoltzmann}, \ref{eqn:Boltzmann}), the free streaming term $- \frac{{\bs k}}{2M} \cdot \nabla_{\bs x}f_{nl}({\bs x}, {\bs k}, t)$ comes from the von-Newmann evolution of the density matrix in the Schr\"odinger picture, i.e., $-i[H_S+\sum_{a,b}\sigma_{ab}L_{ab}, \rho_S]$. This has been explained in Ref.~\cite{Yao:2018nmy} in detail, and will not be repeated here. We will explain the dissociation $\ml{C}_{nl}^-$ and recombination $\ml{C}_{nl}^+$ collision terms in the following. In the derivation of $\ml{C}_{nl}^+$, a semiclassical expansion will also be used.

\subsection{Dissociation}
\label{sect:disso}
\begin{figure}[h]
    \begin{subfigure}[t]{0.49\textwidth}
        \centering
        \includegraphics[height=2.4in]{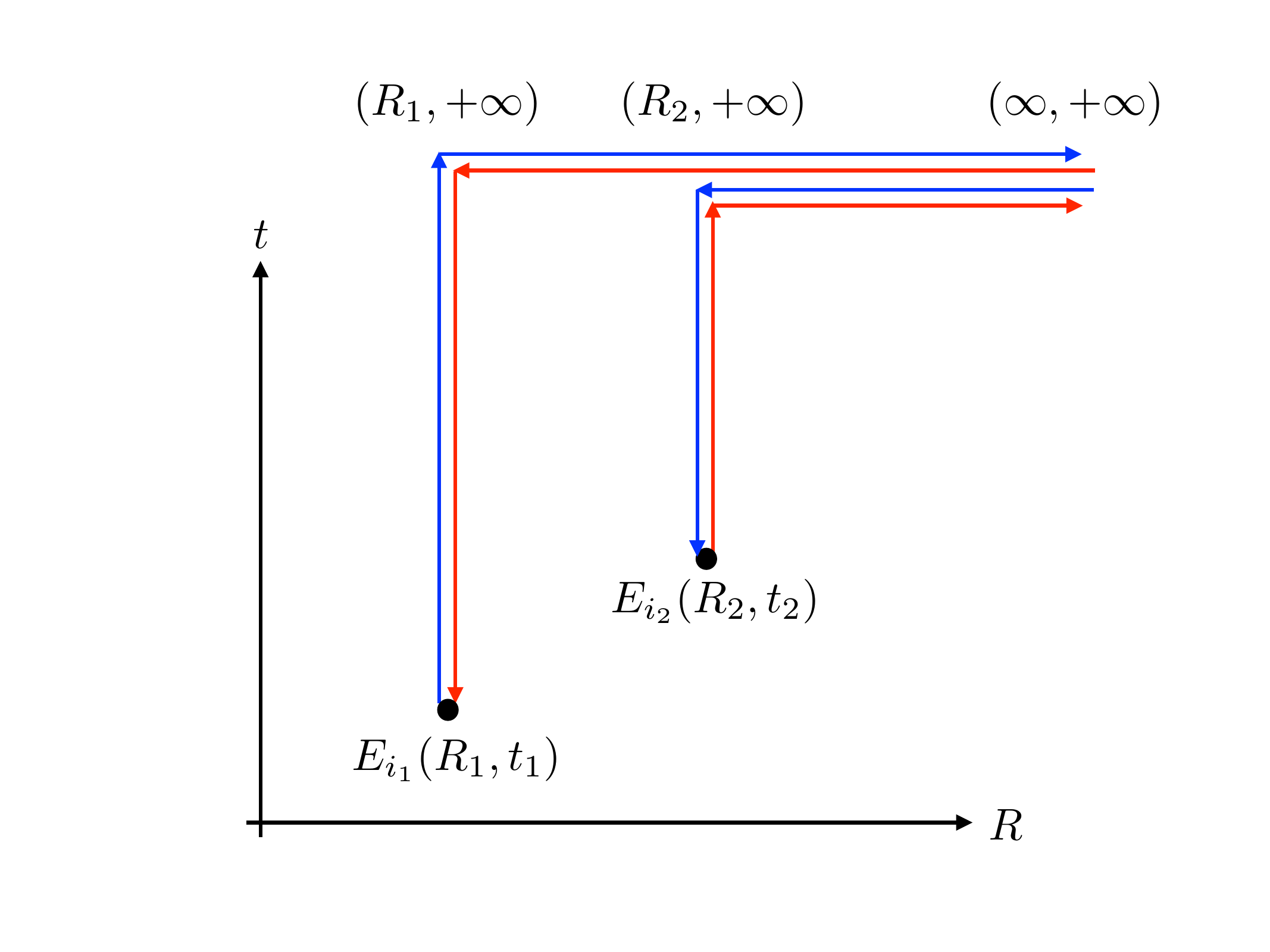}
        \caption{For dissociation, $g^{E++}_{i_1i_2}$.}
    \end{subfigure}%
    ~
    \begin{subfigure}[t]{0.49\textwidth}
        \centering
        \includegraphics[height=2.4in]{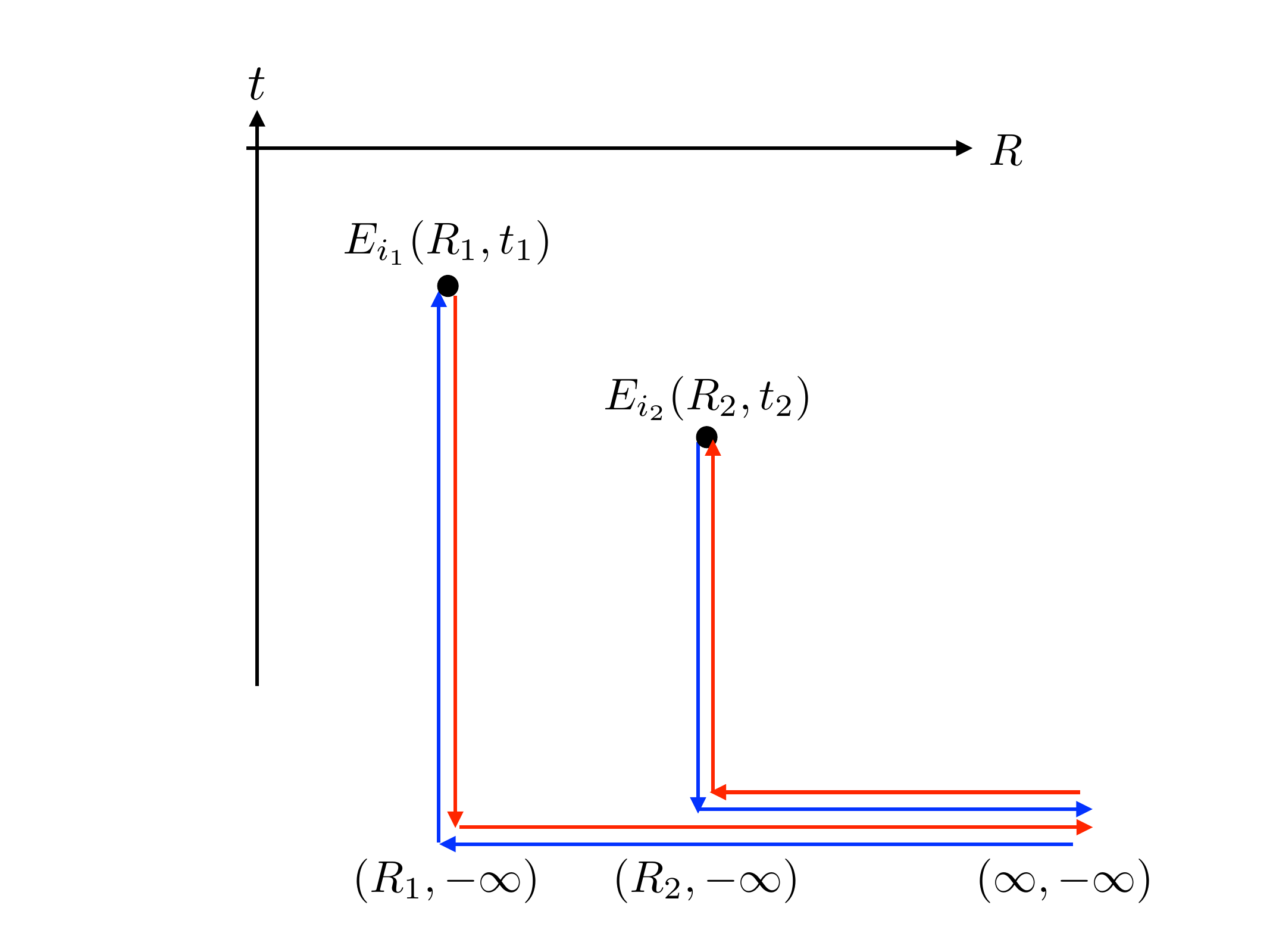}
        \caption{For recombination, $g^{E--}_{i_2i_1}$.}
    \end{subfigure}%
    \caption{Staple-shaped Wilson lines in the definition of the chromoelectric gluon distribution functions of the thermal quark-gluon plasma. The double arrow indicates the adjoint representation. The spatial Wilson lines at infinite time cancel partially and only the part between ${\bs R}_1$ and ${\bs R}_2$ remains. The existence and necessity of the spatial Wilson lines at infinite time are explained in the Appendix~\ref{appendix}.}
    \label{fig:wilson}
\end{figure}

We will first work out the dissociation term $\ml{C}_{nl}^-$ from $\sum_{a,b,c,d}\gamma_{ab,cd}(t) L_{cd}^\dagger L_{ab} \rho_S(-t/2)$. When we sandwich it between $\langle  {\bs k}_1, nl,1 |$ and $| {\bs k}_2, nl, 1 \rangle$, as in the Wigner transform (\ref{eqn:wigner}), we find the state $|d\rangle $ in $L_{cd}^\dagger$ must be $|{\bs k}_1, nl,1 \rangle $ and $|c\rangle = |a \rangle$. Since at the order we are working, the only vertex that couples the color singlet state and the environment is the singlet-octet dipole interaction, we must have $|c\rangle = |a \rangle = | {\bs p}_{\ma{cm}}, {\bs p}_{\ma{rel}},A\rangle(t/2)$ where the $A$ denotes the color of the octet state. By the same argument, we find $|b\rangle = |{\bs k}_3, n'l', 1\rangle$.
So we need to compute
\be 
&&\sum_{a,b,c,d}\gamma_{ab,cd}(t) L_{cd}^\dagger L_{ab} \rho_S \Big(\frac{-t}{2}\Big)
\nn \\ \nn
&=& g^2 \frac{T_F}{N_c} \sum_{n',l'} \sum_{A,B_1,B_2} \sum_{i_1,i_2} \int\frac{\diff^3p_{\ma{cm}}}{(2\pi)^3} \frac{\diff^3p_{\ma{rel}}}{(2\pi)^3} \frac{\diff^3k_{3}}{(2\pi)^3} \int \diff^3R_1 \int \diff^3R_2 \int_{\frac{-t}{2}}^{\frac{t}{2}} \diff t_1 \int_{\frac{-t}{2}}^{\frac{t}{2}} \diff t_2 \\ \nn
&\times& \Tr_E \Big[ \langle {\bs k}_1, nl, 1 | \langle S({\bs R}_1, t_1) | r_{i_1} | \widetilde{O}^{B_1}({\bs R}_1, t_1) \rangle | {\bs p}_{\ma{cm}}, {\bs p}_{\ma{rel}},A\rangle \\ \nn
&\times& \langle {\bs p}_{\ma{cm}}, {\bs p}_{\ma{rel}},A |  \langle \widetilde{O}^{B_2}({\bs R}_2, t_2) | r_{i_2} | S({\bs R}_2, t_2) \rangle | {\bs k}_3, n'l', 1\rangle \\[4pt]
\label{eqn:step_d0}
&\times&  \widetilde{E}_{i_1}^{\dagger B_1}({\bs R}_1, t_1)  \widetilde{E}_{i_2}^{B_2}({\bs R}_2, t_2)  \rho_E  \Big] 
\Big \langle {\bs k}_3, n'l', 1 \Big| \rho_S\Big(\frac{-t}{2}\Big) \Big| {\bs k}_2, nl, 1 \Big\rangle \,.
\ee
Under the Markovian approximation, we can set the upper limits of the two time integrals to infinity $t\to+\infty$.\footnote{The limit $t\to+\infty$ in the time integral and the limit $t\to0$ when we obtained (\ref{eqn:Boltzmann}) from (\ref{eqn:preBoltzmann}) are not contradictory in the Markovian limit. What seems to be a short time to the subsystem during its relaxation, is actually a long time in terms of the environment correlation. The Boltzmann equation is coarse-grained.} Then the octet state $| {\bs p}_{\ma{cm}}, {\bs p}_{\ma{rel}}, A\rangle(t/2)$ can be thought of as an ``asymptotic" outgoing state, which is defined at $t\to +\infty$ by the original octet field creation operator $O^A$. Using Eqs.~(\ref{eqn:O_redef2}, \ref{eqn:O_quanti}, \ref{eqn:s_create}, \ref{eqn:octet_create}), we find
\be
\label{eqn:1}
&&\langle {\bs k}_1, nl, 1 | \langle S({\bs R}_1, t_1) | r_{i_1} |\widetilde{O}^{B_1}({\bs R}_1, t_1) \rangle | {\bs p}_{\ma{cm}}, {\bs p}_{\ma{rel}},A\rangle \nn\\[4pt]
&=& e^{i(E_{nl}t_1 - {\bs k}_1\cdot {\bs R}_1 )} e^{-i(E_pt_1 - {\bs p}_\ma{cm} \cdot {\bs R}_1)} \ml{W}^{B_1A}_{[({\bs R}_1, t_0), ({\bs R}_1, t/2)]} \langle \psi_{nl} | r_{i_1} |\Psi_{{\bs p}_\ma{rel}} \rangle \,,
\ee
where $E_p = ({\bs p}_\ma{rel})^2/M$.
Similarly,
\be 
\label{eqn:2}
&&\langle {\bs p}_{\ma{cm}}, {\bs p}_{\ma{rel}}, A |  \langle \widetilde{O}^{B_2}({\bs R}_2, t_2) |  r_{i_2} | S({\bs R}_2, t_2) \rangle | {\bs k}_3, n'l', 1\rangle \nn\\[4pt]
&=& e^{i(E_pt_2 - {\bs p}_\ma{cm} \cdot {\bs R}_2)}
e^{-i(E_{n'l'}t_2 - {\bs k}_3\cdot {\bs R}_2 )} \ml{W}^{AB_2}_{[({\bs R}_2, t/2), ({\bs R}_2, t_0)]}  \langle \Psi_{{\bs p}_\ma{rel}} |r_{i_2} | \psi_{n'l'} \rangle \,.
\ee
Plugging everything into Eq.~(\ref{eqn:step_d0}) gives
\be
\nn
&& g^2 \frac{T_F}{N_c} \sum_{n',l'} \sum_{A,B_1,B_2} \sum_{i_1,i_2} \int\frac{\diff^3p_{\ma{cm}}}{(2\pi)^3} \frac{\diff^3p_{\ma{rel}}}{(2\pi)^3} \frac{\diff^3k_{3}}{(2\pi)^3} \int \diff^3R_1 \int \diff^3R_2 \int_{\frac{-t}{2}}^{\frac{t}{2}} \diff t_1 \int_{\frac{-t}{2}}^{\frac{t}{2}} \diff t_2 \\ \nn
&\times&   e^{i(E_{nl}t_1 - {\bs k}_1\cdot {\bs R}_1 )  -i(E_pt_1 - {\bs p}_\ma{cm} \cdot {\bs R}_1) } 
       e^{-i(E_{n'l'}t_2 - {\bs k}_3\cdot {\bs R}_2 )  + i(E_pt_2 - {\bs p}_\ma{cm} \cdot {\bs R}_2) } 
        \\ \nn
&\times& \Tr_E\Big[ \ml{W}^{B_1A}_{[({\bs R}_1, t_0), ({\bs R}_1, t/2)]}  \ml{W}^{AB_2}_{[({\bs R}_2, t/2), ({\bs R}_2, t_0)]} 
\widetilde{E}_{i_1}^{\dagger B_1}({\bs R}_1, t_1)  \widetilde{E}_{i_2}^{B_2}({\bs R}_2, t_2)  \rho_E  \Big] \\
\label{eqn:step_d1}
&\times&
\langle \psi_{nl} | r_{i_1} | \Psi_{{\bs p}_\ma{rel}} \rangle  \langle \Psi_{{\bs p}_\ma{rel}} | r_{i_2} | \psi_{n'l'} \rangle
 \Big\langle {\bs k}_3, n'l', 1 \Big| \rho_S\Big(\frac{-t}{2}\Big) \Big| {\bs k}_2, nl, 1 \Big\rangle  \,.
\ee
Together with the redefined chromoelectric fields (\ref{eqn:E},~\ref{eqn:Edagger}), the term inside the partial trace over the environment degrees of freedom can be written as (in the Markovian limit $t\to +\infty$)
\be
&&\lim_{t\to+\infty} \sum_{A,B_1,B_2,C_1,C_2}  \Tr_E \Big[ \ml{W}^{B_1A}_{[({\bs R}_1, t_0), ({\bs R}_1, t/2)]}  \ml{W}^{AB_2}_{[({\bs R}_2, t/2), ({\bs R}_2, t_0)]}
 \nn\\
&\times& E^{C_1}_{i_1}({\bs R}_1,t_1) \ml{W}^{C_1B_1}_{[({\bs R}_1, t_1),({\bs R}_1,t_0)]}
\ml{W}^{B_2C_2}_{[({\bs R}_2, t_0),({\bs R}_2,t_2)]}  E^{C_2}_{i_2}({\bs R}_2,t_2) \rho_E \Big]\nn\\
&=& \frac{1}{T_F} \Big\langle \Tr_{\text{color}} \pig( E_{i_1}({\bs R}_1,t_1) \ml{W}_{[({\bs R}_1, t_1), ({\bs R}_1, +\infty)]}   
\ml{W}_{[({\bs R}_2, +\infty),({\bs R}_2,t_2)]} E_{i_2}({\bs R}_2,t_2) \pig)
\Big\rangle_T \nn\\
&\equiv& \frac{1}{T_F} g^{E++}_{i_1i_2}(t_1, t_2, {\bs R}_1, {\bs R}_2) \,,
\ee
%\be
%&&\sum_{\overline{ij}} \sum_{ \overline{m_1n_1}} \sum_{ \overline{m_2n_2}} \sum_{ \overline{k_1\ell_1}} \sum_{ \overline{k_2\ell_2}} \Tr_E \Big[ W^{im_1}_{[({\bs R}_1, t), ({\bs R}_1, t_R)]} W^{n_1 j}_{[({\bs R}_1, t_L), ({\bs R}_1, t)]}  W^{ n_2 i }_{[({\bs R}_2, t_R),({\bs R}_2, t)]}  W_{[({\bs R}_2, t),({\bs R}_2, t_L)]}^{ jm_2 }  \nn \\ 
%&\times& W^{m_1k_1}_{[({\bs R}_1, t_R),({\bs R}_1,t_1)]} E_{i_1}^{\overline{k_1 \ell_1}}({\bs R}_1,t_1)  W^{\ell_1 n_1}_{[({\bs R}_1, t_1),({\bs R}_1,t_L)]}  \nn\\
%&\times& W^{m_2k_2}_{[({\bs R}_2, t_L),({\bs R}_2,t_2)]}  E^{\overline{k_2\ell_2}}_{i_2}({\bs R}_2,t_2) W^{\ell_2n_2}_{[({\bs R}_2,t_2),({\bs R}_2, t_R)]} \rho_E \Big] \nn \\ \nn
%&\equiv& \frac{1}{T_F} \Big\langle W_{[({\bs R}_1, \infty), ({\bs R}_1, t_1)]} E_{i_1}({\bs R}_1, t_1) W_{[({\bs R}_1, t_1), ({\bs R}_1, \infty)]}  \nn\\
%&\times& W_{[({\bs R}_2, \infty), ({\bs R}_2, t_2)]}  E_{i_2}({\bs R}_2, t_2) W_{[({\bs R}_2, t_2), ({\bs R}_2, \infty)]} \Big\rangle_T \nn \\
%&\equiv& g^{E++}_{i_1i_2}({\bs R}_1, {\bs R}_2, t_1, t_2)\,,
%\ee
where $E_i = E_i^A T^A$ and $\langle \cdots \rangle_T = \Tr_E(\cdots \rho_E)$. In the last line we defined the chromoelectric gluon distribution function $g^{E++}_{i_1i_2}$ of the thermal QGP. 
In the definition of $g^{E++}_{i_1i_2}$, the chromoelectric fields at different spacetime points are dressed with timelike Wilson 
lines extending to infinity. A spatial Wilson line connecting the open ends of the timelike Wilson lines is needed to restore gauge invariance, however, it cannot be generated from the field redefinition applied above. The spatial Wilson line at the infinite time is generated from resumming offshell Coulomb modes which are exchanged between the color octet pair and the medium. Details of the calculation are shown in Appendix~\ref{appendix}. The timelike and spatial Wilson lines together form a staple shape, as shown in Fig.~\ref{fig:wilson}. The shape of the Wilson lines is similar to the case of the gluon TMDPDF, though the time direction here is along the real time, rather than the lightcone time.
%We believe the Wilson line at $t=+\infty$ is only essential in the temporal axial gauge, according to the argument given in Ref.~\cite{Belitsky:2002sm}. The Wilson line at $t=+\infty$ can be neglected in other gauges.

If we assume the thermal QGP is invariant under spacetime translation, we can define
\be
\label{eqn:FT_g++}
&&g^{E++}_{i_1i_2}(t_1, t_2, {\bs R}_1, {\bs R}_2) = g^{E++}_{i_1i_2}( t_1-t_2, {\bs R}_1-{\bs R}_2) \nn\\
&\equiv& \int\frac{\diff^4q}{(2\pi)^4} e^{i q^0(t_1-t_2) - i {\bs q}\cdot ({\bs R}_1-{\bs R}_2) } g^{E++}_{i_1i_2}(q^0, {\bs q}) \,.\ \ \ 
\ee
We can plug the Fourier transform (\ref{eqn:FT_g++}) into (\ref{eqn:step_d1}) and carry out the integrals over ${\bs R}_i$ and $t_i$ where $i=1$, $2$. The spatial integrals lead to $(2\pi)^6\delta^3({\bs k_1} - {\bs p}_\ma{cm} + {\bs q})
\delta^3({\bs k_3} - {\bs p}_\ma{cm} + {\bs q})$. The two time integrals in the Markovian limit $t\to+\infty$ give $\propto \delta(E_{nl}-E_p+q^0)\delta(E_{n'l'}-E_p+q^0)$. If we assume quarkonium states with different quantum numbers $nl$ are non-degenerate (they have different binding energies), then the summation over $n'l'$ will fix $n'=n$ and $l'=l$. Finally using the following integral
\be
\int_{\frac{-t}{2}}^{\frac{t}{2}} \diff t_1 \int_{\frac{-t}{2}}^{\frac{t}{2}} \diff t_2\, e^{i\omega t_1}e^{-i\omega t_2} \xrightarrow{t\to+\infty} 2\pi t\delta(\omega) \,,
\ee
we find (\ref{eqn:step_d1}) can be simplified to
\be
\label{eqn:step_d2}
&& t g^2\frac{1}{N_c} \sum_{i_1,i_2} \int\frac{\diff^3p_{\ma{cm}}}{(2\pi)^3} \frac{\diff^3p_{\ma{rel}}}{(2\pi)^3} \frac{\diff^3k_{3}}{(2\pi)^3} \frac{\diff^4q}{(2\pi)^4} (2\pi)^7\delta^3({\bs k_1} - {\bs p}_\ma{cm} +{\bs q})
\delta^3({\bs k_3} - {\bs p}_\ma{cm} + {\bs q}) \\
&\times& \delta(E_{nl}-E_p+q^0)  
\langle \psi_{nl} | r_{i_1} | \Psi_{{\bs p}_\ma{rel}} \rangle 
\langle \Psi_{{\bs p}_\ma{rel}} | r_{i_2} | \psi_{nl} \rangle 
g^{E++}_{i_1i_2}(q^0,{\bs q})
\langle {\bs k}_3, nl, 1| \rho_S(-t/2) | {\bs k}_2, nl, 1\rangle \,.\nn
\ee
Defining the quarkonium dipole transition function
\be
d_{i_1i_2}^{nl}({\bs p_{\ma{rel}}}) \equiv g^2\frac{1}{N_c}\langle \psi_{nl} | r_{i_1} | \Psi_{{\bs p}_\ma{rel}} \rangle 
\langle \Psi_{{\bs p}_\ma{rel}} | r_{i_2} | \psi_{nl} \rangle \,,
\ee
and making the Wigner transform (\ref{eqn:wigner}) (by setting ${\bs k}_1={\bs k}+\frac{{\bs k}'}{2}$, ${\bs k}_2={\bs k}-\frac{{\bs k}'}{2}$ and shifting ${\bs p}_\ma{cm} \to {\bs p}_\ma{cm} + \frac{{\bs k}'}{2} $), we find (\ref{eqn:step_d2}) turns to
\be
&& t \sum_{i_1,i_2} \int\frac{\diff^3p_{\ma{cm}}}{(2\pi)^3} \frac{\diff^3p_{\ma{rel}}}{(2\pi)^3} \frac{\diff^4q}{(2\pi)^4} (2\pi)^4\delta^3({\bs k} - {\bs p}_\ma{cm} + {\bs q}) \delta(E_{nl}-E_p + q^0) \nn\\
\label{eqn:step_d3}
&\times& d_{i_1i_2}^{nl}({\bs p_{\ma{rel}}}) g^{E++}_{i_1i_2}(q^0,{\bs q})
f_{nl}({\bs x}, {\bs k}, -t/2) \equiv t \ml{C}_{nl}^-({\bs x}, {\bs k}, -t/2) \,.
\ee
So far, we only consider the $\sum_{a,b,c,d}\gamma_{ab,cd}(t) L_{cd}^\dagger L_{ab} \rho_S(-t/2)$ term in the Lindblad equation (\ref{eqn:lindblad}). The other term in the anti-commutator $\sum_{a,b,c,d}\gamma_{ab,cd}(t) \rho_S(-t/2) L_{cd}^\dagger L_{ab} $ can be shown to give the same result (\ref{eqn:step_d3}). Their sum will cancel the factor of $\frac{1}{2}$ in Eq.~(\ref{eqn:lindblad}). So in the Markovian limit, after the Wigner transform, the anti-commutator term in the Lindblad equation leads to $t\ml{C}_{nl}^-({\bs x}, {\bs p}, -t/2)$, as previously shown in Eq.~(\ref{eqn:preBoltzmann}). It should be pointed out that in the derivation of the dissociation collision term, no semiclassical approximation is made.

\subsection{Recombination}
To derive the recombination term $\ml{C}_{nl}^+$ from the Lindblad equation (\ref{eqn:lindblad}), we need to sandwich $\sum_{a,b,c,d}\gamma_{ab,cd}(t)L_{ab} \rho_S(-t/2) L_{cd}^\dagger$ between $\langle  {\bs k}_1, nl,1 |$ and $| {\bs k}_2, nl, 1 \rangle$ and apply the Wigner transform (\ref{eqn:wigner}). We find the state $|a\rangle $ in $L_{ab}$ is $|{\bs k}_1, nl,1 \rangle $ and $|c\rangle$ in $L_{cd}^\dagger$ is $|{\bs k}_2, nl,1 \rangle$. Since at the order we are working, the only vertex that couples the color singlet and the environment is the singlet-octet dipole interaction, we must have $|b \rangle = | {\bs p}_{1\ma{cm}}, {\bs p}_{1\ma{rel}}, A_1\rangle$ and $|d\rangle = | {\bs p}_{2\ma{cm}}, {\bs p}_{2\ma{rel}}, A_2\rangle$ where the $A_i$ denotes the color of the octet state. We need to compute
\be
&& \sum_{a,b,c,d}\gamma_{ab,cd}(t)L_{ab} \rho_S\Big(\frac{-t}{2}\Big) L_{cd}^\dagger \nn\\
&=& g^2 \frac{T_F}{N_c} \sum_{A_1,A_2} \sum_{B_1,B_2} \sum_{i_1,i_2} \int\frac{\diff^3p_{1\ma{cm}}}{(2\pi)^3} \frac{\diff^3p_{1\ma{rel}}}{(2\pi)^3} 
\frac{\diff^3p_{2\ma{cm}}}{(2\pi)^3} \frac{\diff^3p_{2\ma{rel}}}{(2\pi)^3} 
\int \diff^3 R_1 \int \diff^3R_2 \int_{\frac{-t}{2}}^{\frac{t}{2}} \diff t_1 \int_{\frac{-t}{2}}^{\frac{t}{2}} \diff t_2 \nn \\
&\times& \Tr_E \Big[ \langle {\bs k}_1, nl, 1 | \langle S({\bs R}_1, t_1) | r_{i_1} | \widetilde{O}^{B_1}({\bs R}_1, t_1) \rangle | {\bs p}_{1\ma{cm}}, {\bs p}_{1\ma{rel}}, A_1\rangle \nn \\ 
&\times& \langle {\bs p}_{2\ma{cm}}, {\bs p}_{2\ma{rel}},A_2 |  \langle \widetilde{O}^{B_2}({\bs R}_2, t_2) | r_{i_2} | S({\bs R}_2, t_2) \rangle | {\bs k}_2, nl, 1\rangle \nn\\ 
\label{eqn:step_r0}
&\times&    \widetilde{E}_{i_2}^{B_2}({\bs R}_2, t_2) \widetilde{E}_{i_1}^{\dagger B_1}({\bs R}_1, t_1) \rho_E  \Big] 
\Big\langle {\bs p}_{1\ma{cm}}, {\bs p}_{1\ma{rel}}, A_1 \Big| \rho_S\Big(\frac{-t}{2}\Big) \Big| {\bs p}_{2\ma{cm}}, {\bs p}_{2\ma{rel}}, A_2 \Big\rangle \,.
\ee
Now we note that the octet states $| {\bs p}_{i\,\ma{cm}}, {\bs p}_{i\,\ma{rel}}, A_i \rangle$ ($i=1,2$) are defined at $-t/2$ by the original octet field $O^{A_i}$, since they sandwich $\rho_S(-t/2)$. Similar to Eqs.~(\ref{eqn:1}, \ref{eqn:2}), we can show
\be
&&\langle {\bs k}_1, nl, 1 | \langle S({\bs R}_1, t_1) | r_{i_1} |\widetilde{O}^{B_1}({\bs R}_1, t_1) \rangle | {\bs p}_{1\ma{cm}}, {\bs p}_{1\ma{rel}},A_1\rangle  \nn\\[4pt]
&=& e^{i(E_{nl}t_1 - {\bs k}_1\cdot {\bs R}_1 )} e^{-i(E_{p_1}t_1 - {\bs p}_{1\ma{cm}} \cdot {\bs R}_1)} \ml{W}^{B_1A_1}_{[({\bs R}_1, t_0), ({\bs R}_1, -t/2)]} \langle \psi_{nl} | r_{i_1} |\Psi_{{\bs p}_{1\ma{rel}}} \rangle \\[4pt]
%= |\widetilde{O}^{\overline{m_1n_1}}({\bs R}_1, t_1) \rangle | {\bs p}_{1\ma{cm}}, {\bs p}_{1\ma{rel}}, \overline{e_1f_1}\rangle \nn\\[4pt]
%&=& e^{-i(E_{p_1}t_1 - {\bs p}_{1\ma{cm}} \cdot {\bs R}_1)} W^{e_1m_1}_{[({\bs R}_1, 0), ({\bs R}_1, t_R)]} W^{n_1 f_1}_{[({\bs R}_1, t_L), ({\bs R}_1, 0)]}  |\Psi_{{\bs p}_{1\ma{rel}}} \rangle  \\[4pt] 
&&\langle {\bs p}_{2\ma{cm}}, {\bs p}_{2\ma{rel}}, A_2 |  \langle \widetilde{O}^{B_2}({\bs R}_2, t_2) |  r_{i_2} | S({\bs R}_2, t_2) \rangle | {\bs k}_2, nl, 1\rangle \nn\\[4pt]
&=&
e^{i(E_{p_2}t_2 - {\bs p}_{2\ma{cm}} \cdot {\bs R}_2)} 
e^{-i(E_{nl}t_2 - {\bs k}_2\cdot {\bs R}_2 )}
\ml{W}^{A_2B_2}_{[({\bs R}_2, -t/2), ({\bs R}_2, t_0)]}  \langle \Psi_{{\bs p}_{2\ma{rel}}} | r_{i_2} | \psi_{nl} \rangle\,,
%= \langle {\bs p}_{2\ma{cm}}, {\bs p}_{2\ma{rel}}, \overline{e_2f_2} |  \langle \widetilde{O}^{\overline{m_2n_2}}({\bs R}_2, t_2) | \nn\\[4pt]
%&=& e^{i(E_{p_2}t_2 - {\bs p}_{2\ma{cm}} \cdot {\bs R}_2)} W^{ n_2 e_2 }_{[({\bs R}_2, t_R),({\bs R}_2, 0)]}  W_{[({\bs R}_2, 0),({\bs R}_2, t_L)]}^{ f_2m_2 }  \langle \Psi_{{\bs p}_{2\ma{rel}}} | \,,
\ee
where $E_{p_i} = (\bs p_{i\,\ma{rel}})^2/M$. Plugging into (\ref{eqn:step_r0}) and using the Wilson lines dressed on the chromoelectric fields gives
\be
\label{eqn:step_r1}
&& \sum_{a,b,c,d}\gamma_{ab,cd}(t)L_{ab} \rho_S\Big(\frac{-t}{2}\Big) L_{cd}^\dagger
= g^2 \frac{1}{N_c} \sum_{A_1,A_2} \sum_{i_1,i_2} 
\int\frac{\diff^3p_{1\ma{cm}}}{(2\pi)^3} \frac{\diff^3p_{1\ma{rel}}}{(2\pi)^3} 
\frac{\diff^3p_{2\ma{cm}}}{(2\pi)^3} \frac{\diff^3p_{2\ma{rel}}}{(2\pi)^3}  \nn\\
&\times&
\int \diff^3 R_1 \int \diff^3R_2 \int_{\frac{-t}{2}}^{\frac{t}{2}} \diff t_1 \int_{\frac{-t}{2}}^{\frac{t}{2}} \diff t_2 \,
\langle \psi_{nl} | r_{i_1} | \Psi_{{\bs p}_{1\ma{rel}}} \rangle  
 \langle \Psi_{{\bs p}_{2\ma{rel}}} | r_{i_2} | \psi_{nl} \rangle  \nn \\
&\times&   e^{i(E_{nl}t_1 - {\bs k}_1\cdot {\bs R}_1 )  -i(E_{p_1}t_1 - {\bs p}_{1\ma{cm}} \cdot {\bs R}_1) } 
       e^{-i(E_{nl}t_2 - {\bs k}_2\cdot {\bs R}_2 )  + i(E_{p_2}t_2 - {\bs p}_{2\ma{cm}} \cdot {\bs R}_2) }  \nn\\
 &\times&
\big[ g^{E--}_{i_2i_1}(t_2, t_1, {\bs R}_2, {\bs R}_1) \big]^{A_2A_1}
\Big\langle {\bs p}_{1\ma{cm}}, {\bs p}_{1\ma{rel}}, A_1 \Big| \rho_S\Big(\frac{-t}{2}\Big) \Big| {\bs p}_{2\ma{cm}}, {\bs p}_{2\ma{rel}}, A_2 \Big\rangle\,,  \ \ \ \ \ \ 
%&=& g^2 \frac{T_F}{N_c} \sum_{\overline{e_1f_1}} \sum_{\overline{e_2f_2}} \sum_{i_1,i_2} \int\frac{\diff^3p_{1\ma{cm}}}{(2\pi)^3} \frac{\diff^3p_{1\ma{rel}}}{(2\pi)^3} \frac{\diff^3p_{2\ma{cm}}}{(2\pi)^3} \frac{\diff^3p_{2\ma{rel}}}{(2\pi)^3} \int \diff^3 R_1 \int \diff^3R_2 \int_0^t \diff t_1 \int_0^t \diff t_2 \nn \\ 
%&\times&   e^{i(E_{nl}t_1 - {\bs k}_1\cdot {\bs R}_1 )  -i(E_{p_1}t_1 - {\bs p}_{1\ma{cm}} \cdot {\bs R}_1) } e^{-i(E_{nl}t_2 - {\bs k}_2\cdot {\bs R}_2 )  + i(E_{p_2}t_2 - {\bs p}_{2\ma{cm}} \cdot {\bs R}_2) } \langle \psi_{nl} | r_{i_1} | \Psi_{{\bs p}_{1\ma{rel}}} \rangle  \nn \\ 
%&\times& \langle \Psi_{{\bs p}_{2\ma{rel}}} | r_{i_2} | \psi_{nl} \rangle \big[ g^{E--}_{i_1i_2}({\bs R}_2, {\bs R}_1, t_2, t_1) \big]^{\overline{f_2e_2},\overline{e_1f_1}} \Big\langle {\bs p}_{1\ma{cm}}, {\bs p}_{1\ma{rel}}, \overline{e_1f_1} \Big| \rho_S(0) \Big| {\bs p}_{2\ma{cm}}, {\bs p}_{2\ma{rel}}, \overline{e_2f_2} \Big\rangle \nn \\
\ee
where the function $\big[ g^{E--}_{i_2i_1}( t_2, t_1, {\bs R}_2, {\bs R}_1) \big]^{A_2A_1}$ is defined in the Markovian limit by
\be
\label{eqn:pre_g--}
&&\big[ g^{E--}_{i_2i_1}(t_2, t_1, {\bs R}_2, {\bs R}_1) \big]^{A_2A_1} \equiv T_F\lim_{t\to+\infty} \sum_{B_1,B_2,C_1,C_2}\Tr_E\Big[ \ml{W}^{B_1A_1}_{[({\bs R}_1, t_0), ({\bs R}_1, -t/2)]}  \nn\\
&\times& \ml{W}^{A_2B_2}_{[({\bs R}_2, -t/2), ({\bs R}_2, t_0)]}
\ml{W}^{B_2C_2}_{[({\bs R}_2, t_0),({\bs R}_2,t_2)]} 
E^{C_2}_{i_2}({\bs R}_2,t_2)
E^{C_1}_{i_1}({\bs R}_1,t_1) \ml{W}^{C_1B_1}_{[({\bs R}_1, t_1),({\bs R}_1,t_0)]}\rho_E  \Big] \nn\\
&=& T_F\Big\langle  \pig(\ml{W}_{[({\bs R}_2, -\infty),({\bs R}_2,t_2)]} E_{i_2}({\bs R}_2,t_2) \pig)^{A_2}  
\pig(E_{i_1}({\bs R}_1,t_1) \ml{W}_{[({\bs R}_1, t_1),({\bs R}_1,-\infty)]}  \pig)^{A_1}
\Big\rangle_T \,.
%&&\big[ g^{E--}_{i_1i_2}({\bs R}_2, {\bs R}_1, t_2, t_1) \big]^{\overline{f_2e_2},\overline{e_1f_1}} \nn\\
%&\equiv &\sum_{\overline{m_1n_1}} \sum_{\overline{m_2n_2}} \sum_{\overline{k_1\ell_1}} \sum_{\overline{k_2\ell_2}}\Tr_E\Big[ W^{e_1m_1}_{[({\bs R}_1, 0), ({\bs R}_1, t_R)]} W^{n_1 f_1}_{[({\bs R}_1, t_L), ({\bs R}_1, 0)]}  W^{ n_2 e_2 }_{[({\bs R}_2, t_R),({\bs R}_2, 0)]}  W_{[({\bs R}_2, 0),({\bs R}_2, t_L)]}^{ f_2m_2 }  \nn \\ 
%&\times& W^{m_2k_2}_{[({\bs R}_2, t_L),({\bs R}_2,t_2)]}  E^{\overline{k_2\ell_2}}_{i_2}({\bs R}_2,t_2) W^{\ell_2n_2}_{[({\bs R}_2,t_2),({\bs R}_2, t_R)]}   \nn\\
%&\times& W^{m_1k_1}_{[({\bs R}_1, t_R),({\bs R}_1,t_1)]} E_{i_1}^{\overline{k_1 \ell_1}}({\bs R}_1,t_1)  W^{\ell_1 n_1}_{[({\bs R}_1, t_1),({\bs R}_1,t_L)]} \rho_E  \Big] \nn\\
%&\equiv& \Big\langle \big[ W_{[({\bs R}_2, 0),({\bs R}_2,t_2)]}  E_{i_2}({\bs R}_2,t_2) W_{[({\bs R}_2,t_2),({\bs R}_2, 0)] } \big]^{\overline{f_2e_2}}  \nn \\
%&\times& \big[W_{[({\bs R}_1, 0),({\bs R}_1,t_1)]} E_{i_1}({\bs R}_1,t_1) W_{[({\bs R}_1, t_1),({\bs R}_1,0)]} \big]^{\overline{e_1f_1}} \Big\rangle_T \,.
\ee
The newly defined function $(g^{E--}_{i_2i_1})^{A_2A_1}$ is different from the previously defined chromoelectric gluon distribution of the thermal QGP in two aspects: First,  $(g^{E--}_{i_2i_1})^{A_2A_1}$ has open color indexes and thus one may worry that it is gauge dependent. However, the timelike Wilson line together with the spatial Wilson which is explained in Appendix~\ref{appendix} connects the chromoelectric field to a point at ${\bs R} = \bs\infty, t=-\infty$. So the combination $\ml{W}_{[(\bs\infty,-\infty),({\bs R}, t)]} E({\bs R}, t)$ transforms as an adjoint representation at ${\bs R} = \bs\infty, t=-\infty$, which can be set to be a trivial transformation by a choice of the global gauge. Therefore, both $\ml{W}E$ and $E\ml{W}$ are gauge invariant objects independently in the definition of $(g^{E--}_{i_2i_1})^{A_2A_1}$. Second, the end point of the Wilson lines along the time axis is $t=-\infty$ in $(g^{E--}_{i_2i_1})^{A_2A_1}$ rather than $t=+\infty$. The physical interpretation of the different end points of the time axis is as follows: In quarkonium dissociation, the color octet state is a final state and the Wilson line resums the $A_0$ interaction between the octet state and the medium, which is not suppressed by the nonrelativistic expansion. Since only final state interactions are involved, the Wilson lines go to $t=+\infty$. In quarkonium recombination, the color octet state is an initial state and the Wilson line resums the $A_0$ interaction before recombination occurs, which is an initial state interaction.
%In the Markovian limit, the state $\langle {\bs p}_{1\ma{cm}}, {\bs p}_{1\ma{rel}}, A_1 | \rho_S(-t/2) | {\bs p}_{2\ma{cm}}, {\bs p}_{2\ma{rel}}, A_2 \rangle$ can be thought of as an ``asymptotic" incoming state.
But one should keep in mind that the density matrix for the incoming state may be off-diagonal in color space and still contribute to recombination.

%One can think of $t=-\infty$ as the time when the thermal medium is formed, before which no thermal medium exists. Thus, we can extend the time end point $t=0$ to $-\infty$ in the definition of $(g^{E--}_{i_1i_2})^{A_2A_1}$. 

Recombination from the state with an off-diagonal color density matrix is genuinely a quantum effect with no classical analog.
%, since one cannot define a Wigner transform (\ref{eqn:wigner}) for discrete quantum numbers such as the color. 
To further simplify the recombination term (\ref{eqn:step_r1}) and derive the recombination term in the Boltzmann equation, we need to make a semiclassical approximation, which will be explained in the next subsection.

\subsection{Semiclassical Approximation in Recombination}
\label{sect:semi}
As discussed above, we will make semiclassical approximation to write the recombination term (\ref{eqn:step_r1}) as a collision term in the Boltzmann equation (\ref{eqn:Boltzmann}). First we approximate the subsystem density matrix by its diagonal piece in the color space
\be
\label{eqn:color_semi}
 \Big\langle {\bs p}_{1\ma{cm}}, {\bs p}_{1\ma{rel}}, A_1 \Big| \rho_S\Big(\frac{-t}{2}\Big) \Big| {\bs p}_{2\ma{cm}}, {\bs p}_{2\ma{rel}}, A_2 \Big\rangle
\approx \delta^{A_1A_2}
\Big\langle {\bs p}_{1\ma{cm}}, {\bs p}_{1\ma{rel}} \Big| \rho^{(8)}_S\Big(\frac{-t}{2}\Big) \Big| {\bs p}_{2\ma{cm}}, {\bs p}_{2\ma{rel}} \Big\rangle \,,
\ \ \ \ 
\ee
where the superscript $(8)$ indicates the density matrix is a color octet state.
With this approximation, we can contract the color indexes in $(g^{E--}_{i_2i_1})^{A_2A_1}$ and define
\be
&& g^{E--}_{i_2i_1}(t_2, t_1, {\bs R}_2, {\bs R}_1) \equiv \sum_{A_1,A_2}  \delta^{A_1A_2} \big[ g^{E--}_{i_2i_1}(t_2, t_1, {\bs R}_2, {\bs R}_1) \big]^{A_2A_1} \nn\\
&=& T_F \Big\langle \sum_A  \pig(\ml{W}_{[({\bs R}_2, -\infty),({\bs R}_2,t_2)]} E_{i_2}({\bs R}_2,t_2) \pig)^{A} 
\pig(E_{i_1}({\bs R}_1,t_1) \ml{W}_{[({\bs R}_1, t_1),({\bs R}_1,-\infty)]}  \pig)^{A}
\Big\rangle_T \nn\\
&=&  \Big\langle \Tr_{\text{color}} \pig(\ml{W}_{[({\bs R}_2, -\infty),({\bs R}_2,t_2)]} E_{i_2}({\bs R}_2,t_2) 
E_{i_1}({\bs R}_1,t_1) \ml{W}_{[({\bs R}_1, t_1),({\bs R}_1,-\infty)]}  \pig)
\Big\rangle_T\,.
\ee
The function $g^{E--}_{i_2i_1}(t_2, t_1, {\bs R}_2, {\bs R}_1)$ is another chromoelectric gluon distribution function of the thermal QGP, similar to the previously defined $g^{E++}_{i_1i_2}(t_1, t_2, {\bs R}_1, {\bs R}_2)$. The only difference is the orientation of the Wilson line. For $g^{E++}_{i_1i_2}(t_1, t_2, {\bs R}_1, {\bs R}_2)$, the Wilson line goes to $t\to +\infty$ while for $g^{E--}_{i_2i_1}(t_2, t_1, {\bs R}_2, {\bs R}_1)$, the Wilson line comes from $t\to-\infty$. 
%We have extended the time end point $t=0$ in (\ref{eqn:pre_g--}) to $-\infty$ since no medium exists before $t=0$, as discussed in the previous subsection.
A spatial Wilson line connecting the end points of the timelike Wilson lines is also needed for gauge invariance and can be generated from resumming Coulomb modes, as in the case of $g^{E++}$. Detailed calculations of the spatial Wilson line at infinite time can be found in Appendix~\ref{appendix}. The timelike and spatial Wilson lines in the definition of $g^{E--}_{i_2i_1}(t_2, t_1, {\bs R}_2, {\bs R}_1)$ form a staple shape, as plotted in Fig.~\ref{fig:wilson}.

Using the assumption of translational invariance, we have
\be
\label{eqn:FT_g--}
&&g^{E--}_{i_2i_1}(t_2, t_1, {\bs R}_2, {\bs R}_1) = g^{E--}_{i_2i_1}( t_2-t_1, {\bs R}_2-{\bs R}_1) \nn\\
&\equiv& \int\frac{\diff^4q}{(2\pi)^4} e^{i q^0(t_2-t_1) - i {\bs q}\cdot ({\bs R}_2-{\bs R}_1) } g^{E--}_{i_2i_1}(q^0, {\bs q}) \,.\ \ \ 
\ee
Plugging everything into (\ref{eqn:step_r1}) and integrating over ${\bs R}_1$ and ${\bs R}_2$ leads to
\be
\label{eqn:step_r2}
&& g^2\frac{1}{N_c}\sum_{i_1,i_2}
\int\frac{\diff^3p_{1\ma{cm}}}{(2\pi)^3} \frac{\diff^3p_{1\ma{rel}}}{(2\pi)^3} 
\frac{\diff^3p_{2\ma{cm}}}{(2\pi)^3} \frac{\diff^3p_{2\ma{rel}}}{(2\pi)^3}
\frac{\diff^4q}{(2\pi)^4}
\int_{\frac{-t}{2}}^{\frac{t}{2}} \diff t_1 \int_{\frac{-t}{2}}^{\frac{t}{2}} \diff t_2  \nn \\ 
&\times&  e^{i(E_{nl}-E_{p_1}-q^0)t_1} e^{-i(E_{nl}-E_{p_2}-q^0)t_2} 
(2\pi)^6\delta^3({\bs k}_1 - {\bs p}_{1\ma{cm}} - {\bs q}) \delta^3({\bs k}_2 - {\bs p}_{2\ma{cm}} - {\bs q}) \nn \\[4pt]
&\times&  \langle \psi_{nl} | r_{i_1} | \Psi_{{\bs p}_{1\ma{rel}}} \rangle
\langle \Psi_{{\bs p}_{2\ma{rel}}} | r_{i_2} | \psi_{nl} \rangle
 g^{E--}_{i_2i_1}(q^0,{\bs q}) \Big\langle {\bs p}_{1\ma{cm}}, {\bs p}_{1\ma{rel}} \Big| \rho^{(8)}_S\Big(\frac{-t}{2}\Big) \Big| {\bs p}_{2\ma{cm}}, {\bs p}_{2\ma{rel}} \Big\rangle \,.\ \ \ \ \ 
\ee 
When applying the Wigner transform to (\ref{eqn:step_r2}), we set ${\bs k}_1 = {\bs k} + {\bs k}'/2$ and ${\bs k}_2 = {\bs k} - {\bs k}'/2$. Changing variables ${\bs p}_{1\ma{cm}} \to {\bs p}_{\ma{cm}} + {\bs p}'_{\ma{cm}}/2$ and ${\bs p}_{2\ma{cm}} \to {\bs p}_{\ma{cm}} - {\bs p}'_{\ma{cm}}/2$, we find
\be
\label{eqn:step_r3}
&& g^2\frac{1}{N_c}\sum_{i_1,i_2}
\int\frac{\diff^3p_{\ma{cm}}}{(2\pi)^3} \frac{\diff^3p_{1\ma{rel}}}{(2\pi)^3} 
 \frac{\diff^3p_{2\ma{rel}}}{(2\pi)^3}
\frac{\diff^4q}{(2\pi)^4}
\int_{\frac{-t}{2}}^{\frac{t}{2}} \diff t_1 \int_{\frac{-t}{2}}^{\frac{t}{2}} \diff t_2 \, e^{i(E_{nl}-E_{p_1}-q^0)t_1} e^{-i(E_{nl}-E_{p_2}-q^0)t_2} \nn \\
&\times&   
(2\pi)^3\delta^3({\bs k} - {\bs p}_{\ma{cm}} - {\bs q}) 
 \langle \psi_{nl} | r_{i_1} | \Psi_{{\bs p}_{1\ma{rel}}} \rangle
\langle \Psi_{{\bs p}_{2\ma{rel}}} | r_{i_2} | \psi_{nl} \rangle
 g^{E--}_{i_2i_1}(q^0,{\bs q}) \nn\\
 &\times& \Big\langle {\bs p}_{\ma{cm}}+\frac{{\bs k}'}{2}, {\bs p}_{1\ma{rel}} \Big| \rho^{(8)}_S\Big(\frac{-t}{2}\Big) \Big| {\bs p}_{\ma{cm}}-\frac{{\bs k}'}{2}, {\bs p}_{2\ma{rel}} \Big\rangle \,.
\ee
Applying the Wigner transform, we find
\be
\label{eqn:wigner8}
&&\int\frac{\diff^3 k'}{(2\pi)^3} e^{i{\bs k}'\cdot{\bs x}_\ma{cm}} \Big\langle {\bs p}_{\ma{cm}}+\frac{{\bs k}'}{2}, {\bs p}_{1\ma{rel}} \Big| \rho^{(8)}_S\Big(\frac{-t}{2}\Big) \Big| {\bs p}_{\ma{cm}}-\frac{{\bs k}'}{2}, {\bs p}_{2\ma{rel}} \Big\rangle \nn\\
&=& \int \diff^3 x_\ma{rel} e^{-i({\bs p}_{1\ma{rel}} - {\bs p}_{2\ma{rel}}) \cdot {\bs x}_\ma{rel} } f_{Q\bar{Q}}^{(8)} \Big({\bs x}_\ma{cm}, {\bs p}_{\ma{cm}}, {\bs x}_\ma{rel}, \frac{ {\bs p}_{1\ma{rel}}+{\bs p}_{2\ma{rel}} }{2}, \frac{-t}{2} \Big) \,,
\ee
where $f_{Q\bar{Q}}^{(8)} ({\bs x}_\ma{cm}, {\bs p}_{\ma{cm}}, {\bs x}_\ma{rel}, \frac{ {\bs p}_{1\ma{rel}}+{\bs p}_{2\ma{rel}} }{2} , -t/2)$ is the phase space distribution function of a color octet $Q\bar{Q}$ pair with center-of-mass and relative positions and momenta ${\bs x}_\ma{cm}, {\bs p}_{\ma{cm}}, {\bs x}_\ma{rel}, \frac{ {\bs p}_{1\ma{rel}}+{\bs p}_{2\ma{rel}} }{2}$. If the color reaches thermal equilibrium, statistically we will have
\be
f_{Q\bar{Q}}^{(8)} = \frac{N_c^2-1}{N_c^2}f_{Q\bar{Q}} \,,
\ee
where $f_{Q\bar{Q}}$ is the distribution function of an unbound $Q\bar{Q}$ pair that can be either a color singlet state or an octet state.

Now we take a crucial step in the derivation of the semiclassical transport equation: the semiclassical expansion, also known as the gradient expansion. A general discussion of the gradient expansion in the derivation of semiclassical transport equations can be found, for example, in Ref.~\cite{haug2008quantum}. We will expand $f_{Q\bar{Q}}^{(8)}$ around some ${\bs x}_{\ma{rel}} = {\bs x}_0$ and assume the distribution varies slowly as ${\bs x}_{\ma{rel}}$ changes
\be
\label{eqn:gradient}
f_{Q\bar{Q}}^{(8)}({\bs x}_\ma{cm}, {\bs p}_{\ma{cm}}, {\bs x}_\ma{rel}, \frac{ {\bs p}_{1\ma{rel}}+{\bs p}_{2\ma{rel}} }{2}, t ) 
= f_{Q\bar{Q}}^{(8)}({\bs x}_\ma{cm}, {\bs p}_{\ma{cm}}, {\bs x}_0, \frac{ {\bs p}_{1\ma{rel}}+{\bs p}_{2\ma{rel}} }{2} ,t ) \nn\\
+ ( {\bs x}_\ma{rel} - {\bs x}_0) \cdot \nabla_{{\bs x}_0} f_{Q\bar{Q}}^{(8)}({\bs x}_\ma{cm}, {\bs p}_{\ma{cm}}, {\bs x}_0, \frac{ {\bs p}_{1\ma{rel}}+{\bs p}_{2\ma{rel}} }{2},t ) + \cdots \,,
\ee
where higher order terms in the gradient expansion are omitted. In this section, we will focus on the leading term in the gradient expansion. The next-to-leading term, which will be discussed in Section~\ref{sect:quantum},  corresponds to a quantum correction to the semiclassical Boltzmann equation. In practice, we want to choose ${\bs x}_0$ such that 
quantum corrections are minimized. We will choose ${\bs x}_0 = 0$ when we compute the correction in the next section.

The derivation shown in Ref.~\cite{Yao:2018nmy} uses the gradient expansion implicitly, by assuming the distribution function of a $Q\bar{Q}$ pair is uniform in the relative position. This assumption of uniformity is exactly the leading term in the gradient expansion (\ref{eqn:gradient}). The argument given in Ref.~\cite{Yao:2018nmy} relies on a large diffusion rate for open heavy quarks and the angular dependence of the octet state wavefunctions $|\Psi_{{\bs p}_\ma{rel}}\rangle$ (see Eq.~(D8) of Ref.~\cite{Yao:2018nmy}). Thus it is not obvious how to generalize the derivation in Ref.~\cite{Yao:2018nmy} to incorporate quantum corrections. Here the derivation is based on a gradient expansion and higher order corrections can be worked out systematically.

Plugging the leading term in the gradient expansion back into (\ref{eqn:wigner8}), we find the integral over ${\bs x}_\ma{rel}$ can be done trivially which gives $(2\pi)^3\delta^3({\bs p}_{1\ma{rel}} - {\bs p}_{2\ma{rel}})$. Now we can carry out the time integrals in (\ref{eqn:step_r3}) in the Markovian limit when ${\bs p}_{1\ma{rel}} = {\bs p}_{2\ma{rel}} \equiv {\bs p}_{\ma{rel}}$. The time integrals in the Markovian limit have been explained in Section~\ref{sect:disso}. We can show after the Wigner transform (\ref{eqn:wigner8}), Eq.~(\ref{eqn:step_r3}) turns to
\be
\label{eqn:step_r4}
&& t g^2\frac{1}{N_c}\sum_{i_1,i_2}
\int\frac{\diff^3p_{\ma{cm}}}{(2\pi)^3} \frac{\diff^3p_{\ma{rel}}}{(2\pi)^3} 
\frac{\diff^4q}{(2\pi)^4} 
(2\pi)^4\delta^3({\bs k} - {\bs p}_{\ma{cm}} - {\bs q}) 
\delta(E_{nl} - E_p - q^0)  \nn\\
&\times& \langle \psi_{nl} | r_{i_1} | \Psi_{{\bs p}_{\ma{rel}}} \rangle
\langle \Psi_{{\bs p}_{\ma{rel}}} | r_{i_2} | \psi_{nl} \rangle
 g^{E--}_{i_2i_1}(q^0,{\bs q}) f_{Q\bar{Q}}^{(8)}({\bs x}_\ma{cm}, {\bs p}_{\ma{cm}}, {\bs x}_0, {\bs p}_{\ma{rel}}, -t/2 ) \nn\\
&=& t \sum_{i_1,i_2}
\int\frac{\diff^3p_{\ma{cm}}}{(2\pi)^3} \frac{\diff^3p_{\ma{rel}}}{(2\pi)^3}
\frac{\diff^4q}{(2\pi)^4} 
(2\pi)^4\delta^3({\bs k} - {\bs p}_{\ma{cm}} - {\bs q}) 
\delta(E_{nl} - E_p - q^0)  \nn\\
&\times& d_{i_1i_2}^{nl}({\bs p}_\ma{rel})
 g^{E--}_{i_2i_1}(q^0,{\bs q}) f_{Q\bar{Q}}^{(8)}({\bs x}_\ma{cm}, {\bs p}_{\ma{cm}}, {\bs x}_0, {\bs p}_{\ma{rel}}, -t/2 ) \equiv t \ml{C}_{nl}^+({\bs x}_\ma{cm}, {\bs k}, -t/2) \,,
\ee
where $E_p = ({\bs p}_\ma{rel})^2/M$ and we defined the collision term for recombination $\ml{C}_{nl}^+$ in the Boltzmann equation (\ref{eqn:Boltzmann}). The structure of (\ref{eqn:step_r4}) is very similar to that of (\ref{eqn:step_d3}). We will discuss these two collision terms in the next subsection.

\subsection{Factorization of Reaction Rates}
In the previous subsections, we have derived the collision terms in the semiclassical Boltzmann equation for dissociation and recombination. From Eqs.~(\ref{eqn:step_d3}, \ref{eqn:step_r4}), we have
\be
\label{eqn:disso}
\ml{C}_{nl}^-({\bs x}, {\bs k}, t) &=& 
\sum_{i_1,i_2} \int\frac{\diff^3p_{\ma{cm}}}{(2\pi)^3} \frac{\diff^3p_{\ma{rel}}}{(2\pi)^3} \frac{\diff^4q}{(2\pi)^4} (2\pi)^4\delta^3({\bs k} - {\bs p}_\ma{cm} + {\bs q}) \delta(E_{nl}-E_p + q^0) \nn\\
&\times& d_{i_1i_2}^{nl}({\bs p_{\ma{rel}}}) g^{E++}_{i_1i_2}(q^0,{\bs q})   f_{nl}({\bs x}, {\bs k}, t) \\
\label{eqn:reco}
\ml{C}_{nl}^+({\bs x}_\ma{cm}, {\bs k}, t) &=& \sum_{i_1,i_2}
\int\frac{\diff^3p_{\ma{cm}}}{(2\pi)^3} \frac{\diff^3p_{\ma{rel}}}{(2\pi)^3}
\frac{\diff^4q}{(2\pi)^4} 
(2\pi)^4\delta^3({\bs k} - {\bs p}_{\ma{cm}} - {\bs q}) 
\delta(E_{nl} - E_p - q^0)  \nn\\
&\times& d_{i_1i_2}^{nl}({\bs p}_\ma{rel})
 g^{E--}_{i_2i_1}(q^0,{\bs q}) f_{Q\bar{Q}}^{(8)}({\bs x}_\ma{cm}, {\bs p}_{\ma{cm}}, {\bs x}_0, {\bs p}_{\ma{rel}}, t ) \,.
\ee
Then the dissociation and recombination rates can be defined. General expressions for the reaction rates can be found in Refs.~\cite{Yao:2018zze,Yao:2018sgn}. We will first study the dissociation rate of a quarkonium state with position ${\bs x}$ and momentum ${\bs k}$, which can be written as
\be
R^{-}_{nl}({\bs x}, {\bs k}, t ) = \frac{\ml{C}_{nl}^-({\bs x}, {\bs k}, t)}{f_{nl}({\bs x}, {\bs k}, t)} \,,
\ee
where the rate depends on position and time via the dependence of the QGP temperature on position and time.\footnote{We assume the spacetime variation of the QGP temperature is much slower than the typical relaxation time of quarkonium, such that during quarkonium dissociation or recombination, the QGP can be treated as translationally invariant in spacetime.} The QGP temperature determines the chromoelectric gluon distribution function. Using (\ref{eqn:disso}), we find
\be
R^{-}_{nl}({\bs x}, {\bs k}, t ) &=& \sum_{i_1,i_2} \int\frac{\diff^3p_{\ma{cm}}}{(2\pi)^3} \frac{\diff^3p_{\ma{rel}}}{(2\pi)^3} \frac{\diff^4q}{(2\pi)^4} (2\pi)^4\delta^3({\bs k} - {\bs p}_\ma{cm} + {\bs q}) \delta(E_{nl}-E_p + q^0) \nn\\
&\times& d_{i_1i_2}^{nl}({\bs p_{\ma{rel}}}) g^{E++}_{i_1i_2}(q^0,{\bs q})  \,.
\ee
The summation over $i_1,i_2$ can be further simplified. So far we have not written out the dependence on the third component of the orbital angular momentum $m_l$ explicitly. In practice, we will average over $m_l$ since current heavy ion experiments do not measure $m_l$. Temporarily restoring the $m_l$ dependence in the bound state wavefunction, we obtain (note that in the integrand, the only dependence on $\hat{\bs p}_{\ma{rel}}$ is in the dipole transition function)
\be
\label{eqn:average}
&&\frac{1}{2l+1}\sum_{m_l=-l}^l \int \diff \Omega_{{\bs p}_\ma{rel}}d_{i_1i_2}^{nl m_l}({\bs p_{\ma{rel}}}) \nn\\ &=&\frac{1}{2l+1}\sum_{m_l=-l}^l \int \diff \Omega_{{\bs p}_\ma{rel}} g^2\frac{1}{N_c} \langle \psi_{nl m_l} | r_{i_1} | \Psi_{{\bs p}_\ma{rel}} \rangle 
\langle \Psi_{{\bs p}_\ma{rel}} | r_{i_2} | \psi_{nl m_l} \rangle \nn\\
&=& \frac{\delta_{i_1i_2}}{3}\frac{1}{2l+1}\sum_{m_l=-l}^l \int \diff \Omega_{{\bs p}_\ma{rel}} g^2\frac{1}{N_c} |\langle \psi_{nl m_l} | {\bs r} | \Psi_{{\bs p}_\ma{rel}} \rangle|^2 \nn\\
&\equiv&  \delta_{i_1i_2} \int \diff \Omega_{{\bs p}_\ma{rel}} \overline{d}_{nl}({\bs p_{\ma{rel}}})
\ee
where $\diff \Omega_{{\bs p}_\ma{rel}} = \diff\cos\theta_{{\bs p}_\ma{rel}} \diff \phi_{{\bs p}_\ma{rel}}$. Defining
\be
g^{E++}(q^0,{\bs q}) \equiv \sum_{i_1,i_2} \delta_{i_1i_2}g^{E++}_{i_1i_2}(q^0,{\bs q}) \,,
\ee
we can write the dissociation rate as
\be
\label{eqn:disso1}
R^{-}_{nl} &=& \int\frac{\diff^3p_{\ma{cm}}}{(2\pi)^3} \frac{\diff^3p_{\ma{rel}}}{(2\pi)^3} \frac{\diff^4q}{(2\pi)^4} (2\pi)^4\delta^3({\bs k} - {\bs p}_\ma{cm} + {\bs q}) \delta(E_{nl}-E_p + q^0) \overline{d}_{nl}({\bs p_{\ma{rel}}}) g^{E++}(q^0,{\bs q})  \nn\\
&=& \int\frac{\diff^3p_{\ma{cm}}}{(2\pi)^3} \frac{\diff^3p_{\ma{rel}}}{(2\pi)^3}
\overline{d}_{nl}({\bs p_{\ma{rel}}}) g^{E++}\Big(\frac{({\bs p}_\ma{rel})^2}{M} - E_{nl},{\bs p}_\ma{cm} - {\bs k} \Big) \nn\\
&=& \int \frac{\diff^3p_{\ma{rel}}}{(2\pi)^3} \overline{d}_{nl}({\bs p_{\ma{rel}}}) G^{E++}\Big(\frac{({\bs p}_\ma{rel})^2}{M} - E_{nl}\Big) \,,
\ee
where $G^{E++}$ is the integrated chromoelectric gluon distribution function:
\be
G^{E++}(q^0) &\equiv& \int\frac{\diff^3 q}{(2\pi)^3} g^{E++}(q^0, {\bs q}) = \int \diff t \, e^{-i q^0t} g^{E++}(t, {\bs 0}) \nn\\
&=& \int \diff t \, e^{-i q^0t}  \Big\langle \Tr_{\text{color}} \pig( E_i(t) \ml{W}_{[t,0]} E_i(0) \pig) \Big\rangle_T \,.
\ee
In the integrated chromoelectric gluon distribution function $G^{E++}$, the spatial index $i$ is summed over and $\ml{W}$ denotes a Wilson line in the adjoint representation. The Wilson lines connecting ${\bs R}_1$ and ${\bs R}_2$ shown in Fig.~\ref{fig:wilson} overlap at the same position and only the part between $t_1$ and $t_2$ remains after cancellation.

The differential rate can be written as
\be
\label{eqn:disso2}
(2\pi)^3 \frac{\diff R^{-}_{nl}}{\diff^3p_{\ma{cm}}} &=& \int \frac{\diff^3p_{\ma{rel}}}{(2\pi)^3} \frac{\diff^4q}{(2\pi)^4} (2\pi)^4\delta^3({\bs k} - {\bs p}_\ma{cm} + {\bs q}) \delta(E_{nl}-E_p+q^0) \overline{d}_{nl}({\bs p_{\ma{rel}}}) g^{E++}(q^0,{\bs q}) \nn\\
&=& \int \frac{\diff^3p_{\ma{rel}}}{(2\pi)^3}  \overline{d}_{nl}({\bs p_{\ma{rel}}}) g^{E++}\Big(\frac{({\bs p}_\ma{rel})^2}{M} - E_{nl},{\bs p}_\ma{cm} - {\bs k} \Big) \,.
\ee
Eqs.~(\ref{eqn:disso1}, \ref{eqn:disso2}) are important results of this paper. They show the inclusive and differential dissociation rates of quarkonium factorize into the quarkonium dipole transition function $\overline{d}_{nl}$ and the chromoelectric gluon distribution function of the QGP. In the inclusive rate, the gluon distribution function is momentum independent while in the differential rate, it is momentum dependent. The connection between the Wilson line structures in the definitions of the momentum independent and momentum dependent chromoelectric gluon distribution functions is very similar to the relation between the gluon PDF and the gluon TMDPDF. Through the use of this factorization theorem, experimental measurements of quarkonium nuclear modification factors probe the chromoelectric gluon distribution function of the QGP. 
The centrality dependence of the quarkonium nuclear modification factor probes the momentum independent distribution while the transverse momentum dependence and measurements of the azimuthal angular anisotropy may be able to probe the momentum dependent distribution, since both the differential dissociation and recombination rates depend on the momentum dependent distribution.

One application of the factorization formula (\ref{eqn:disso1}) is to combine it with lattice QCD calculations to constrain the real part of the in-medium potential of quarkonium. The thermal width extracted from lattice QCD calculations of the spectral functions contains both the dissociation rate and the diffusion rate. In the diffusion process, the singlet $Q\bar{Q}$ pair exchanges some momentum with the medium but does not break up. The diffusion process is suppressed with respect to dissociation when the temperature is small, i.e., $rT \ll 1$ where $r\sim 1/(Mv)$ is the typical size of a quarkonium state. In our power counting, the dissociation amplitude scales as $rT$ while the diffusion amplitude scales as $(rT)^2$ \cite{Yao:2018sgn}. So at leading order in the multipole expansion, the thermal width is equal to the dissociation rate $R^-$. Then if the chromoelectric gluon distribution function $G^{E++}$ can be calculated in lattice QCD, we can combine the two lattice calculations and use the factorization formula (\ref{eqn:disso1}) to constrain the quarkonium dipole transition function $\overline{d}_{nl}$. The dipole transition is between the bound state and the unbound scattering state. Their wavefunctions can be solved by using parametrized in-medium real potentials. So we can calculate $\overline{d}_{nl}$ with different parametrizations of the real potential to compare with the one constrained by lattice QCD calculations. This method can indirectly constrain the in-medium real potential of quarkonium. It may also be used to test the consistency between the real and imaginary parts of the potential calculated in lattice QCD~\cite{Rothkopf:2011db}. In practice, one may first carry out the above analysis for $\Upsilon(1S)$ at low temperature, where the power counting parameter is small and the framework presented here is under good theoretical control. Recent lattice QCD calculations of the thermal width and other developments for bottomonium at finite temperature can be found in Refs.~\cite{Larsen:2019bwy,Larsen:2019zqv,Larsen:2020rjk}.

Next we will study the recombination term $\ml{C}^+$. Using (\ref{eqn:average}) and integrating over ${\bs p}_\ma{cm}$, we find (\ref{eqn:reco}) becomes
\be
\label{eqn:reco1}
&&\ml{C}_{nl}^+({\bs x}_\ma{cm}, {\bs k}, t) \nn\\
&=&
\int \frac{\diff^3p_{\ma{rel}}}{(2\pi)^3}
\frac{\diff^3q}{(2\pi)^3} 
\overline{d}_{nl}({\bs p}_\ma{rel})
 g^{E--}\Big(E_{nl}-\frac{({\bs p}_\ma{rel})^2}{M},{\bs q}\Big) f_{Q\bar{Q}}^{(8)}({\bs x}_\ma{cm}, {\bs k} - {\bs q}, {\bs x}_0, {\bs p}_{\ma{rel}}, t ) \,,
\ee
which factorizes into three pieces: dipole transition function, chromoelectric gluon distribution function and the octet $Q\bar{Q}$ distribution function. Eq.~(\ref{eqn:reco1}) should be thought of as a differential recombination process because the final state momentum, i.e., the quarkonium momentum, ${\bs k}$, is not integrated over. Integrating over ${\bs k}$ leads to
\be
\label{eqn:reco2}
\int\frac{\diff^3 k}{(2\pi)^3}\ml{C}_{nl}^+({\bs x}_\ma{cm}, {\bs k}, t) =
\int \frac{\diff^3p_{\ma{rel}}}{(2\pi)^3} \overline{d}_{nl}({\bs p}_\ma{rel})
G^{E--}\Big(E_{nl}-\frac{({\bs p}_\ma{rel})^2}{M}\Big) n_{Q\bar{Q}}^{(8)}({\bs x}_\ma{cm}, {\bs x}_0, {\bs p}_{\ma{rel}}, t ) \,,\ \ \ \ \ \ 
\ee
where $G^{E--}$ is the integrated chromoelectric gluon distribution function and $n_{Q\bar{Q}}^{(8)}$ is a density. They are given by
\be
G^{E--}(q^0) &\equiv& \int \frac{\diff^3 q}{(2\pi)^3} g^{E--}(q^0, {\bs q})  = \int \diff t e^{-iq^0t} g^{E--}(t, {\bs 0}) \nn\\
&=& \int \diff t e^{-iq^0t} \Big\langle \Tr_{\text{color}} \pig( E_i(t) \ml{W}_{[t,0]} E_i(0) \pig) \Big\rangle_T \\
n_{Q\bar{Q}}^{(8)}({\bs x}_\ma{cm}, {\bs x}_0, {\bs p}_{\ma{rel}}, t ) &\equiv& \int\frac{\diff^3 k}{(2\pi)^3} 
f_{Q\bar{Q}}^{(8)}({\bs x}_\ma{cm}, {\bs k}, {\bs x}_0, {\bs p}_{\ma{rel}}, t ) \,.
\ee
We note that the integrated chromoelectric gluon distributions $G^{E++}$ and $G^{E--}$ are the same.
Converting Eqs.~(\ref{eqn:reco1}, \ref{eqn:reco2}) into differential and inclusive recombination rates of a heavy quark requires knowledge of the relation between the two-particle $Q\bar{Q}$ distribution and the one-particle $\bar{Q}$ distribution. We will not pursue writing out the recombination rates explicitly here. Relevant formulas can be found in Ref.~\cite{Yao:2018sgn}.

Finally, we comment on the scale dependence of each component in the factorization formula. The chromoelectric gluon distribution function has a natural scale $T$, the plasma temperature. So we need to compute the dipole transition function $\overline{d}_{nl}$ at the scale $T$. The EFT pNRQCD is constructed by matching at the scale $Mv$, we need to solve the renormalization group equation for $\overline{d}_{nl}$ from $Mv$ to $T$. It has been shown that at one loop, no extra renormalization is needed for the dipole interaction vertex beyond the renormalization of the strong coupling constant $\alpha_s$ \cite{Pineda:2000gza}. We believe this is true to all orders due to the reparametrization invariance of the Lagrangian $\bar{\psi}(iD_0 - \frac{{\bs D}^2}{2M})\psi$ for a single heavy quark field $\psi$, from which the leading pNRQCD Lagrangian is derived (see the derivation of pNRQCD in \cite{Fleming:2005pd}).

\section{Quantum Correction to Semiclassical Transport}
\label{sect:quantum}
In this section, we work out the leading quantum correction to the semiclassical Boltzmann transport equation. For dissociation, no semiclassical expansion is applied. For recombination, we make two semiclassical approximations. The first one is (\ref{eqn:color_semi}), where we assume the octet state density matrix is diagonal in the color space. We want to point out that this assumption is not necessary if one works at leading order in the coupling constant. The reason why we have to make this semiclassical assumption is the open color indexes in (\ref{eqn:pre_g--}). If we only keep the leading terms (in the coupling constant) in (\ref{eqn:pre_g--}), we can set all the Wilson lines to be unity. Then (\ref{eqn:pre_g--}) becomes
\be
\big\langle E^{A_2}_{i_2}({\bs R}_2, t_2)  E^{A_1}_{i_1}({\bs R}_1, t_1) \big\rangle_T \,,
\ee
which is proportional to $\delta^{A_1A_2}$ up to higher order corrections (in the coupling constant). Therefore, at leading order, only the diagonal entries of the color density matrix contribute to recombination. But in general, off-diagonal entries can contribute.  To derive the recombination collision term in semiclassical Boltzmann equation, we have to approximate the octet density matrix to be diagonal in color.

The second semiclassical approximation is the gradient expansion (\ref{eqn:gradient}). So far, we only take the leading term in the gradient expansion. We now work out the recombination term from the next-to-leading term: $( {\bs x}_\ma{rel} - {\bs x}_0) \cdot \nabla_{{\bs x}_0} f_{Q\bar{Q}}^{(8)}({\bs x}_\ma{cm}, {\bs p}_{\ma{cm}}, {\bs x}_0, \frac{ {\bs p}_{1\ma{rel}}+{\bs p}_{2\ma{rel}} }{2},t )$. For simplicity, we will set ${\bs x}_0=0$. In practice, one wants to choose a ${\bs x}_0$ such that the gradient expansion converges fastest. With the next-to-leading term, (\ref{eqn:wigner8}) becomes
\be
&& \int \diff^3 x_\ma{rel} e^{-i({\bs p}_{1\ma{rel}} - {\bs p}_{2\ma{rel}})\cdot {\bs x}_\ma{rel}} {\bs x}_\ma{rel}  \cdot \nabla_{{\bs x}_0} f_{Q\bar{Q}}^{(8)} \Big({\bs x}_\ma{cm}, {\bs p}_{\ma{cm}}, {\bs x}_0, \frac{ {\bs p}_{1\ma{rel}}+{\bs p}_{2\ma{rel}} }{2}, t \Big) \bigg|_{{\bs x}_0=0} \\
&=& \int \diff^3 x_\ma{rel} \bigg[ i\nabla_{{\bs p}_{1\ma{rel}}}e^{-i({\bs p}_{1\ma{rel}} - {\bs p}_{2\ma{rel}})\cdot {\bs x}_\ma{rel} }\bigg]
\cdot
\bigg[\nabla_{{\bs x}_0} f_{Q\bar{Q}}^{(8)} \Big({\bs x}_\ma{cm}, {\bs p}_{\ma{cm}}, {\bs x}_0, \frac{ {\bs p}_{1\ma{rel}}+{\bs p}_{2\ma{rel}} }{2}, t \Big) \bigg]_{{\bs x}_0=0} \nn\\
&=& \bigg[ i\frac{\nabla_{{\bs p}_{1\ma{rel}}} - \nabla_{{\bs p}_{2\ma{rel}}} }{2} (2\pi)^3\delta^3({\bs p}_{1\ma{rel}}-{\bs p}_{2\ma{rel}}) \bigg]
\cdot
\bigg[\nabla_{{\bs x}_0} f_{Q\bar{Q}}^{(8)} \Big({\bs x}_\ma{cm}, {\bs p}_{\ma{cm}}, {\bs x}_0, \frac{ {\bs p}_{1\ma{rel}}+{\bs p}_{2\ma{rel}} }{2}, t \Big) \bigg]_{{\bs x}_0=0} \,.\nn
\ee
Plugging this into the Wigner transform of Eq.~(\ref{eqn:step_r3}) and integrating ${\bs p}_{1\ma{rel}}$ and ${\bs p}_{2\ma{rel}}$ by parts leads to
\be
&& g^2\frac{1}{N_c}\sum_{i_1,i_2}
\int\frac{\diff^3p_{\ma{cm}}}{(2\pi)^3} \frac{\diff^3p_{1\ma{rel}}}{(2\pi)^3} 
 \frac{\diff^3p_{2\ma{rel}}}{(2\pi)^3}
\frac{\diff^4q}{(2\pi)^4}
\int_{\frac{-t}{2}}^{\frac{t}{2}} \diff t_1 \int_{\frac{-t}{2}}^{\frac{t}{2}} \diff t_2  e^{i(E_{nl}-E_{p_1}-q^0)t_1} e^{-i(E_{nl}-E_{p_2}-q^0)t_2} \nn \\ 
&\times&   
(2\pi)^6\delta^3({\bs k} - {\bs p}_{\ma{cm}} - {\bs q}) 
\delta^3({\bs p}_{1\ma{rel}}-{\bs p}_{2\ma{rel}})
 \langle \psi_{nl} | r_{i_1} | \Psi_{{\bs p}_{1\ma{rel}}} \rangle
\langle \Psi_{{\bs p}_{2\ma{rel}}} | r_{i_2} | \psi_{nl} \rangle
 g^{E--}_{i_2i_1}(q^0,{\bs q}) \nn\\[4pt]
 &\times& \bigg[ \frac{-i\nabla_{{\bs p}_{1\ma{rel}}} \langle \psi_{nl} | r_{i_1} | \Psi_{{\bs p}_{1\ma{rel}}}\rangle}{2 \langle \psi_{nl} | r_{i_1} | \Psi_{{\bs p}_{1\ma{rel}}}\rangle} + \frac{i\nabla_{{\bs p}_{2\ma{rel}}} \langle \Psi_{{\bs p}_{2\ma{rel}}} | r_{i_2} | \psi_{nl} \rangle}{2 \langle \Psi_{{\bs p}_{2\ma{rel}}} | r_{i_2} | \psi_{nl} \rangle}  - \frac{{\bs p}_{1\ma{rel}}}{M}t_1  - \frac{{\bs p}_{2\ma{rel}}}{M}t_2 \bigg]  \nn\\
 &\cdot&
\bigg[\nabla_{{\bs x}_0} f_{Q\bar{Q}}^{(8)} \Big({\bs x}_\ma{cm}, {\bs p}_{\ma{cm}}, {\bs x}_0, \frac{ {\bs p}_{1\ma{rel}}+{\bs p}_{2\ma{rel}} }{2}, t \Big) \bigg]_{{\bs x}_0=0}  \nn\\[4pt]
&=& t \sum_{i_1,i_2}
\int\frac{\diff^3p_{\ma{cm}}}{(2\pi)^3} \frac{\diff^3p_{\ma{rel}}}{(2\pi)^3}
\frac{\diff^4q}{(2\pi)^4} (2\pi)^4 \delta^3({\bs k} - {\bs p}_{\ma{cm}} - {\bs q})  \delta(E_{nl}-E_p - q^0) d_{i_1i_2}^{nl}({\bs p}_{\ma{rel}}) g^{E--}_{i_2i_1}(q^0,{\bs q}) \nn\\
 &\times& \bigg[ \frac{-i\nabla_{{\bs p}_{\ma{rel}}} \langle \psi_{nl} | r_{i_1} | \Psi_{{\bs p}_{\ma{rel}}}\rangle}{ 2 \langle \psi_{nl} | r_{i_1} | \Psi_{{\bs p}_{\ma{rel}}}\rangle} + \frac{i\nabla_{{\bs p}_{\ma{rel}}} \langle \Psi_{{\bs p}_{\ma{rel}}} | r_{i_2} | \psi_{nl} \rangle}{2 \langle \Psi_{{\bs p}_{\ma{rel}}} | r_{i_2} | \psi_{nl} \rangle} \bigg]  \cdot
\bigg[\nabla_{{\bs x}_0} f_{Q\bar{Q}}^{(8)} \Big({\bs x}_\ma{cm}, {\bs p}_{\ma{cm}}, {\bs x}_0, \ {\bs p}_{\ma{rel}}, t \Big) \bigg]_{{\bs x}_0=0}  \nn\\[4pt]
&\equiv& t \ml{Q}^{+}_{nl}({\bs x}_\ma{cm}, {\bs k}, t) \,,
\ee
where we have used the fact
\be
\int_{\frac{-t}{2}}^{\frac{t}{2}} \diff t_1 \int_{\frac{-t}{2}}^{\frac{t}{2}} \diff t_2 (t_1+t_2) e^{i\omega t_1} e^{-i\omega t_2} = 0\,.
\ee
We have derived the leading quantum correction $\ml{Q}^{+}_{nl}$ to the recombination term $\ml{C}^+_{nl}$ in the semiclassical Boltzmann transport equation. Higher order quantum corrections from the gradient expansion can be similarly worked out. The quantum correction is small when the distribution in the relative position between the $Q\bar{Q}$ pair varies slowly.

\section{Conclusions}
\label{sect:conclusion}
In this paper, we derived the semiclassical Boltzmann equation for quarkonium in the thermal QGP by applying pNRQCD and the open quantum systems framework. Under the hierarchy $M\gg Mv \gg Mv^2,\,T,\,\Lambda_{\text{QCD}}$, we worked at leading order in the power counting parameter $v$ and $\frac{T}{Mv}$, which correspond to nonrelativistic and multipole expansion respectively. In our power counting, the interaction vertex between the subsystem (quarkonium) and the environment (thermal QGP) scales as $\frac{T}{Mv}$ and thus is weak. In the weak coupling (between the subsystem and the environment) limit, the total density matrix factorizes into the subsystem density matrix and the environment density matrix. We demonstrated how the Lindblad equation for quarkonium as an open system turns into a Boltzmann equation after taking the Markovian limit and applying a Wigner transform (a Gaussian smearing is required for maintaining positivity). We justified the Markovian approximation using our power counting. The derivation is valid for a strongly coupled QGP at leading power of the nonrelativistic and multipole expansions since we resummed relevant interactions to all orders in the coupling constant at the scale $Mv^2$ and $T$. Reaction rates in the Boltzmann equation factorize into the quarkonium dipole transition function and the chromoelectric gluon distribution function of the QGP. The factorization originates from the factorization of the total density matrix. The structures of the Wilson line in the chromoelectric gluon distribution function are different for inclusive and differential reaction rates. The relation between the two is similar to that between the gluon PDF and the gluon TMDPDF, except that the time axis here is along the real time rather than the lightcone time. In the recombination, we also made semiclassical approximations. One semiclassical approximation assumes the initial state density matrix is diagonal in color space, while the second semiclassical approximation keeps only the lowest term in the gradient expansion. Finally we worked out the leading quantum correction to the semiclassical Boltzmann transport equation by computing the
next-to-leading term in the gradient expansion.

The factorization in the transport equation allows us to use experimental measurements on quarkonium suppression in heavy ion collisions to probe the chromoelectric gluon distribution functions of the QGP. The chromoelectric gluon distribution functions are defined nonperturbatively here so in principle, they can be computed by using lattice QCD or the AdS/CFT correspondence. It would be interesting to investigate how much perturbative and nonperturbative calculations differ for the distribution function. Once the distribution function is determined nonperturbatively, one can combine it with the thermal width of quarkonium extracted from lattice QCD calculations to constrain the quarkonium in-medium real potential indirectly. In practice, one may choose $\Upsilon(1S)$ at low temperature for a well-controlled power counting. Furthermore, the differential reaction rates depend on a new momentum dependent chromoelectric distribution function defined by two electric fields connected via staple-shaped Wilson lines. It will be interesting to explore what other physics processes are sensitive to this correlation in the QGP. Finally, one can implement the quantum correction to the semiclassical transport equation in phenomenological studies and investigate the importance of quantum corrections. The framework developed here can be easily generalized to study quarkonium transport in cold nuclear matter by replacing the thermal QGP density matrix with a density matrix describing cold nuclear matter, which is relevant for quarkonium production in eA collisions in the future Electron-Ion Collider.

\acknowledgments
We thank Harold U. Baranger, Jean-Paul Blaizot, Nora Brambilla, Miguel A. Escobedo, Hong Liu, Krishna Rajagopal, Iain W. Stewart, Antonio Vairo and Hendrik van Hees for inspiring discussions. XY thanks the organizers of the workshop EMMI Rapid Reaction Task Force: Suppression and (re)generation of quarkonium in heavy-ion collisions at the LHC. This material is based upon work supported by the U.S. Department of Energy, Office of Science, Office of Nuclear Physics under grant Contract Numbers DE-SC0011090 and DE-FG02-05ER41368.

\appendix
\section{Gauge Link at Infinite Time}
\label{appendix}
In the main text, the Wilson lines along the time axis arise due to field redefinitions. Here we show how to obtain the Wilson line along the spatial direction at infinite time. The derivation is similar to that of  Ref.~\cite{Belitsky:2002sm} where the gauge link at infinite lightcone time was derived for the TMDPDF. For our purposes here, we first discuss modes that contribute to this gauge link.

The effective theory pNRQCD is based on separation of scales: $M \gg Mv \gg Mv^2 \gtrsim T$. We will distinguish modes by their momentum scaling $p^\mu = (p^0, p^1, p^2, p^3)$. The hard, soft and ultrasoft modes scale as
\be
p^\mu_{\ma{h}} &\sim& M(1,1,1,1) \\
p^\mu_{\ma{s}} &\sim& M(v,v,v,v) \\
p^\mu_{\ma{us}} &\sim& M(v^2,v^2,v^2,v^2) \,. 
\ee
In the derivation of pNRQCD, the hard and soft modes are integrated out, so this theory describes the dynamics of the ultrasoft modes. The Wilson line along the time axis discussed in the main text resums the interaction between the c.m. motion of the octet and the ultrasoft $A_0$ field.

In fact, another mode with a scaling that falls between the soft and ultrasoft modes can influence the ultrasoft dynamics. It has the momentum scaling
\be
p^\mu_{\ma{c}} \sim M(v^2,v,v,v) \,,
\ee
and thus is named the Coulomb mode. The Coulomb modes that mediate the interaction between the $Q\bar{Q}$ pair has been included in the pNRQCD Lagrangian~(\ref{eq:lagr}) as potentials. However, the Coulomb mode that couples the c.m. motion of the octet with the medium has not been considered yet. As we will show below, it is this Coulomb mode that leads to the gauge link at infinite time.

The relevant part of the Lagrangian density is
\be
\int \diff^3 r\,\Tr \bigg( \ma{O}^\dagger({\bs R}, {\bs r}, t) \Big(iD_0 + \frac{{\bs D}_{\bs R}^2}{4M} + \frac{\nabla_{\bs r}^2}{M} - V_o(\bs r) +\cdots \Big) \ma{O}({\bs R}, {\bs r}, t) \bigg)\,,
\ee
where the new term compared with Eq.~(\ref{eq:lagr}) is the c.m. kinetic energy ${\bs D}_{\bs R}^2 /(4M)$. For ultrasoft gauge fields, the c.m. kinetic energy can be omitted since it is at higher order in the $v$ expansion. However, for Coulomb gauge fields, the c.m. kinetic energy is at the same order as $iD_0$, which is at leading order. In the adjoint representation, the c.m. kinetic energy can be written as
\be
&& \int \diff^3 r\, \Tr\bigg( \ma{O}^\dagger({\bs R}, {\bs r}, t)  \frac{\nabla_{\bs R}^2}{4M}  \ma{O}({\bs R}, {\bs r}, t)  - \frac{ig}{4M}\ma{O}^\dagger({\bs R}, {\bs r}, t) \Big( {\bs A}({\bs R},t) \cdot \nabla_{\bs R} \nn\\[4pt]
&+& \nabla_{\bs R}\cdot {\bs A}({\bs R},t) \Big) \ma{O}({\bs R}, {\bs r}, t) 
- \frac{g^2}{4M} \ma{O}^\dagger({\bs R}, {\bs r}, t) {\bs A}^2({\bs R},t)  \ma{O}({\bs R}, {\bs r}, t) \bigg)\,.
\ee

\begin{figure}
\centering
\includegraphics[height=1.8in]{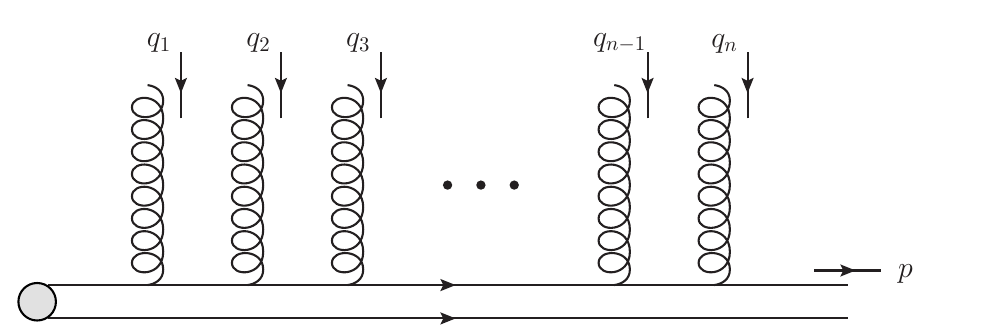}
\caption{A general Feynman diagram in the series contributing to the gauge link at infinite time. The outgoing octet field has ultrasoft momentum $p$. All the attached gluons originate from the interaction term linear in ${\bs A}$ and scale as Coulomb modes.}
\label{fig:coulomb}
\end{figure}

\begin{figure}
\centering
\includegraphics[height=1.8in]{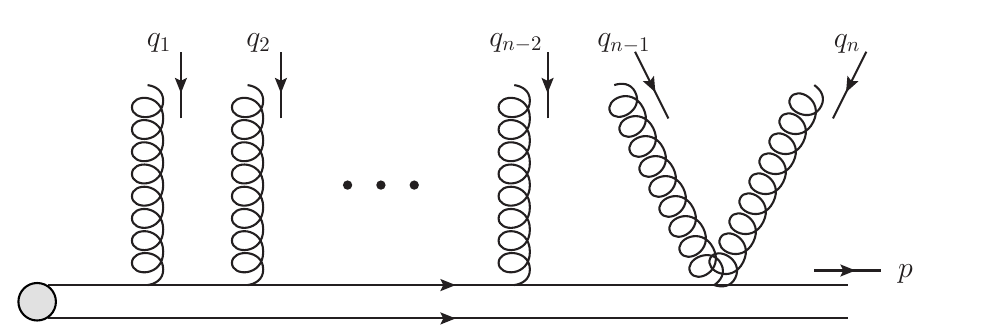}
\caption{Feynman diagram contributing to the gauge link at infinite time, similar to Fig.~\ref{fig:coulomb} except that the rightmost two gauge fields originate from the interaction term quadratic in ${\bs A}$.}
\label{fig:A2}
\end{figure}

We will first focus on the interaction given by the term linear in ${\bs A}$ and discuss the term quadratic in ${\bs A}$ later. For the interaction linear in ${\bs A}$, we will resum a series of Feynman diagrams. The $n$-th diagram is depicted in Fig.~\ref{fig:coulomb}. The outgoing octet field is on-shell and ultrasoft: $p\sim M(v^2,v^2,v^2,v^2)$. All gluon fields interacting with the octet are Coulomb: $q_j \sim M(v^2,v,v,v)$, so ${\bs q}_j + {\bs p} = {\bs q}_j + \ml{O}(Mv^2)$ for the spatial components of the momentum. 
%We can choose a frame by a Galilean transform such that $p = (p^0, {\bs 0})$. 
The amplitude can be written as (we want to keep the ${\bs A}$ fields explicitly for later convenience)
\be
\ml{M}_n&\equiv& \bigg[ \prod_{j=1}^n \int \frac{\diff^4 q_j}{(2\pi)^4} \bigg] e^{-i(-\sum_{j=1}^n{\bs q}_j)\cdot {\bs R}_0} \bigg[ \frac{-g}{4M} \frac{1}{ p^0 - q_n^0 - \frac{{\bs q}_n^2}{4M} + i\epsilon} {\bs q}_n \cdot A(q_n) \bigg] \nn\\
&\times& \bigg[ \frac{-g}{4M} \frac{1}{ p^0 - q_n^0 - q_{n-1}^0 - \frac{({\bs q}_n+{\bs q}_{n-1})^2}{4M} + i\epsilon} \big(2{\bs q}_n+{\bs q}_{n-1} \big) \cdot {\bs A}(q_{n-1}) \bigg] \nn\\ [2pt]
&\times & \ \cdots \nn\\ [2pt] 
\label{eqn:Mn}
&\times& \bigg[ \frac{-g}{4M} \frac{1}{ p^0 -  \sum_{j=1}^n q_j^0 - \frac{(\sum_{j=1}^n{\bs q}_j)^2}{4M} + i\epsilon } \Big(2\sum_{j=2}^{n}{\bs q}_j + {\bs q}_1 \Big) \cdot {\bs A}(q_1) 
\bigg] \,.
\ee
Here ${\bs R}_0$ is the starting position of the octet field sitting on the left end of Fig.~\ref{fig:coulomb}. We set ${\bs R}_0=0$ for simplicity for the moment and discuss the case with ${\bs R}_0\neq0$ later. (We can also keep $t_0$ in the phase where $t_0$ is the starting time of the octet field. But $t_0$ will become irrelevant in the following derivation.) Shifting $q_n\to q_n + p$ and applying the trick used in Ref.~\cite{Belitsky:2002sm}, we find
\be
\frac{1}{- 4Mq_n^0 - {\bs q}_n^2 + i\epsilon}  = -i \int_0^\infty \diff\lambda_n \exp\Big\{i\lambda_n \big[ -4Mq_n^0 - {\bs q}_n^2 + i\epsilon \big] \Big\}\,,
\ee
for the first propagator and a similar expression for a generic propagator 
\be
&&\frac{1}{ - 4M \sum_{j=1}^n q_j^0 - (\sum_{j=1}^n{\bs q}_j)^2 + i\epsilon } \nn\\
&=& -i \int_0^\infty \diff\lambda_1 \exp\Big\{i\lambda_1 \big[
 - 4M \sum_{j=1}^n q_j^0 - (\sum_{j=1}^n{\bs q}_j)^2 + i\epsilon \big] \Big\} \,.
\ee
Then the integral over $q_m^0$ can be thought of as a Fourier transform, with the conjugated time $4M \sum_{j=1}^m \lambda_j$, since the gluon lines are incoming with the phase defined by $e^{-iq \cdot x}$. In the large mass limit, the conjugated time $4M \sum_{j=1}^m \lambda_j \to +\infty$, since the $\lambda_j$'s are positive. The octet field is outgoing in Fig.~\ref{fig:coulomb} so the diagram corresponds to final-state interactions in dissociation. If the octet field is incoming, as in recombination, a different sign will be obtained in the conjugated time, which leads to $-\infty$ in the large mass limit. After the Fourier transform in the large mass limit, the expression (\ref{eqn:Mn}) becomes
\be
\ml{M}_n &=& g^n \bigg[\prod_{j=1}^n \int \frac{\diff^3 q_j}{(2\pi)^3} \bigg] 
\bigg[\frac{{\bs q}_n \cdot {\bs A}(t=+\infty, {\bs q}_n)}{({\bs q}_n)^2 -i\epsilon} \bigg]
\bigg[ \frac{(2{\bs q}_n + {\bs q}_{n-1}) \cdot {\bs A}(t=+\infty, {\bs q}_{n-1})}{({\bs q}_n+{\bs q}_{n-1})^2 -i\epsilon} \bigg]  \nn\\
\label{eqn:Mn2}
&\times& \cdots \times \bigg[ \frac{(2\sum_{j=2}^{n}{\bs q}_j + {\bs q}_1 ) \cdot {\bs A}(t=+\infty, {\bs q}_1)} {(\sum_{j=1}^n {\bs q}_j)^2-i\epsilon}  \bigg] \,.
\ee
Plugging the Fourier transform,
\be
\label{eqn:fourier}
{\bs A}(t=+\infty, {\bs q}_{m}) = \int \diff^3R_m\, e^{-i{\bs q}_m\cdot {\bs R}_m} {\bs A}(t=+\infty, {\bs R}_{m}) \,,
\ee
 into (\ref{eqn:Mn2}) leads to
\be
\ml{M}_n &=& g^n \bigg[\prod_{j=1}^n \int \frac{\diff^3 q_j}{(2\pi)^3} \diff^3R_j \, e^{-i{\bs q}_j\cdot {\bs R}_j} \bigg] 
\bigg[\frac{{\bs q}_n \cdot {\bs A}(t=+\infty, {\bs R}_n)}{({\bs q}_n)^2 -i\epsilon} \bigg] \nn\\
&\times& \bigg[ \frac{2({\bs q}_n + {\bs q}_{n-1}) \cdot {\bs A}(t=+\infty, {\bs R}_{n-1}) + i\nabla_{n-1} \cdot {\bs A}(t=+\infty, {\bs R}_{n-1}) }{({\bs q}_n+{\bs q}_{n-1})^2 -i\epsilon} \bigg] \times  \cdots \nn\\ 
&\times&   \bigg[ \frac{2(\sum_{j=1}^{n}{\bs q}_j  ) \cdot {\bs A}(t=+\infty, {\bs R}_1) + i\nabla_1 \cdot {\bs A}(t=+\infty, {\bs R}_1) } {(\sum_{j=1}^n {\bs q}_j)^2-i\epsilon} \bigg] \,. 
\ee
One can show
\be
\label{eqn:phase_trick}
\prod_{j=1}^n e^{-i{\bs q}_j\cdot {\bs R}_j}
&=& e^{-i{\bs q}_n \cdot ({\bs R}_n - {\bs R}_{n-1})}
e^{-i({\bs q}_n+{\bs q}_{n-1}) \cdot ({\bs R}_{n-1} - {\bs R}_{n-2})} \times \cdots \nn\\
&\times& e^{-i(\sum_{j=2}^n{\bs q}_j) \cdot ({\bs R}_2 - {\bs R}_1)}
e^{-i(\sum_{j=1}^n{\bs q}_j) \cdot {\bs R}_1} \,.
\ee
Then we can do a change of variables, ${\bs k}_m = \sum_{j=m}^n {\bs q}_j $ to simplify the momentum integrals:
\be
\ml{M}_n &=& g^n \bigg[\prod_{j=1}^n \int \diff^3R_j \bigg] \bigg[ {\bs A}(t=+\infty, {\bs R}_n) \cdot i \nabla_n \int \frac{\diff^3 k_n}{(2\pi)^3} \frac{e^{-i{\bs k}_n \cdot ({\bs R}_n - {\bs R}_{n-1})}}{{\bs k}_n^2 - i\epsilon} \bigg] \nn\\
&\times& \bigg[ 2{\bs A}(t=+\infty, {\bs R}_{n-1}) \cdot i \nabla_{n-1} \int \frac{\diff^3 k_{n-1}}{(2\pi)^3} \frac{e^{-i{\bs k}_{n-1} \cdot ({\bs R}_{n-1} - {\bs R}_{n-2})}}{{\bs k}_{n-1}^2 - i\epsilon} \nn\\
&& + \int \frac{\diff^3 k_{n-1}}{(2\pi)^3} \frac{e^{-i{\bs k}_{n-1} \cdot ({\bs R}_{n-1} - {\bs R}_{n-2})}}{{\bs k}_{n-1}^2 - i\epsilon} \, i\nabla_{n-1}\cdot {\bs A}(t=+\infty, {\bs R}_{n-1}) \bigg] \times \cdots \nn\\
&\times& \bigg[ 2{\bs A}(t=+\infty, {\bs R}_{1}) \cdot i \nabla_{1} \int \frac{\diff^3 k_{1}}{(2\pi)^3} \frac{e^{-i{\bs k}_{1} \cdot{\bs R}_{1}}}{{\bs k}_{1}^2 - i\epsilon} \nn\\
&& + \int \frac{\diff^3 k_{1}}{(2\pi)^3} \frac{e^{-i{\bs k}_{1} \cdot {\bs R}_{1}}}{{\bs k}_{1}^2 - i\epsilon} \, i\nabla_{1}\cdot {\bs A}(t=+\infty, {\bs R}_{1}) \bigg] \,.
\ee
Using the standard integral in the derivation of Coulomb potential from the propagator, we find
\be
\ml{M}_n &=& (ig)^n \bigg[\prod_{j=1}^n \int \diff^3R_j \bigg] \bigg[ {\bs A}(t=+\infty, {\bs R}_n) \cdot  \nabla_n \frac{1}{4\pi |{\bs R}_n - {\bs R}_{n-1}| }\bigg] \nn\\
&\times& \bigg[ 2{\bs A}(t=+\infty, {\bs R}_{n-1}) \cdot \nabla_{n-1} \frac{1}{4\pi |{\bs R}_{n-1} - {\bs R}_{n-2}| } \nn\\
&& + \frac{1}{4\pi |{\bs R}_{n-1} - {\bs R}_{n-2}|}  \nabla_{n-1}\cdot {\bs A}(t=+\infty, {\bs R}_{n-1}) \bigg] \times \cdots \nn\\
&\times& \bigg[ 2{\bs A}(t=+\infty, {\bs R}_{1}) \cdot \nabla_{1} \frac{1}{4\pi |{\bs R}_1|} + \frac{1}{4\pi |{\bs R}_1|} \nabla_{1}\cdot {\bs A}(t=+\infty, {\bs R}_{1}) \bigg] \,.
\ee
At time $t=+\infty$, the gauge field is a pure gauge\footnote{This follows from Ref.~\cite{Belitsky:2002sm}. A pure gauge field is given by $A^\mu(x) = \frac{i}{g}U(x) \partial^\mu U^\dagger(x)$ where $U(x)$ is a gauge transformation. We consider a gauge transformation that is a perturbation of the identity: $U(x) \approx 1 + ig\phi^\dagger(x)$ where $\phi^\dagger(x)$ is group valued.}:
\be
\label{eqn:pure_gauge}
{\bs A}(t=+\infty, {\bs R}_m) = \nabla_m \phi( {\bs R}_m ) \,.
\ee
The term with ${\bs A}({\bs R}_n)$ integrated over ${\bs R}_n$ can be written as (we use integration by parts)
\be
ig \int \diff^3 R_n \big[\nabla_n \phi( {\bs R}_n )\big]  \nabla_n \frac{1}{4\pi |{\bs R}_n - {\bs R}_{n-1}|} 
%= \int \diff^3 R_n \phi( {\bs R}_n ) \delta^3({\bs R}_n - {\bs R}_{n-1}) 
= ig\, \phi( {\bs R}_{n-1} )\,.
\ee
Now we keep the term with ${\bs A}({\bs R}_n)$ and the term with ${\bs A}({\bs R}_{n-1})$, then after the integration over ${\bs R}_n$, we find
\be
&& (ig)^2\int \diff^3 R_{n-1} 
\bigg[ 2 \phi( {\bs R}_{n-1} ) {\bs A}(t=+\infty, {\bs R}_{n-1}) \cdot \nabla_{n-1} \frac{1}{4\pi |{\bs R}_{n-1} - {\bs R}_{n-2}| } \nn\\
&& + \phi( {\bs R}_{n-1} ) \frac{1}{4\pi |{\bs R}_{n-1} - {\bs R}_{n-2}|}  \nabla_{n-1}\cdot {\bs A}(t=+\infty, {\bs R}_{n-1}) \bigg] \nn\\[4pt]
&=& (ig)^2\int \diff^3 R_{n-1} 
\bigg[  \phi( {\bs R}_{n-1} ) {\bs A}(t=+\infty, {\bs R}_{n-1}) \cdot \nabla_{n-1} \frac{1}{4\pi |{\bs R}_{n-1} - {\bs R}_{n-2}| } \nn\\
&& - \nabla_{n-1}\phi( {\bs R}_{n-1} ) \frac{1}{4\pi |{\bs R}_{n-1} - {\bs R}_{n-2}|}  \cdot {\bs A}(t=+\infty, {\bs R}_{n-1}) \bigg] \nn\\[4pt]
&=& (ig)^2 \int \diff^3 R_{n-1}\, \phi( {\bs R}_{n-1} ) \big[\nabla_{n-1} \phi({\bs R}_{n-1}) \big] \cdot \nabla_{n-1} \frac{1}{4\pi |{\bs R}_{n-1} - {\bs R}_{n-2}| } \nn\\
&& - (ig)^2 \int \diff^3 R_{n-1} \frac{{\bs A}^2(t=+\infty, {\bs R}_{n-1})}{4\pi |{\bs R}_{n-1} - {\bs R}_{n-2}| } \nn\\[4pt]
\label{eqn:second_order}
&=& \frac{(ig)^2}{2} \phi^2( {\bs R}_{n-2} ) - (ig)^2 \int \diff^3 R_{n-1} \frac{{\bs A}^2(t=+\infty, {\bs R}_{n-1})}{4\pi |{\bs R}_{n-1} - {\bs R}_{n-2}| } \,,
\ee
where we have used $\phi( {\bs R}_{n-1} ) \nabla_{n-1} \phi({\bs R}_{n-1}) = \frac{1}{2} \nabla_{n-1} \phi^2({\bs R}_{n-1})$.
As we will show later, the term $(ig\phi)^2/2$ contributes to the Wilson line structure we are seeking for. The second term in (\ref{eqn:second_order}) will be cancelled exactly by the interaction term quadratic in ${\bs A}$. No analogous cancellation appears  in the derivation of Ref.~\cite{Belitsky:2002sm}. The relevant diagram is shown in Fig.~\ref{fig:A2}. By similar calculations as above, one can show the contribution to the amplitude from the part on the left of the $q_{n-2}$ gluon line in the diagram is the same. The diagram in Fig.~\ref{fig:A2} gives:
\be
&&\bigg[ \prod_{j=1}^n \int \frac{\diff^4 q_j}{(2\pi)^4} \bigg] \bigg[ \frac{-ig^2}{4M} \frac{i}{p^0 - q_{n}^0 - q_{n-1}^0 - \frac{({\bs q}_n + {\bs q}_{n-1} )^2}{4M} +i\epsilon} {\bs A}(q_{n}) \cdot  {\bs A}(q_{n-1}) \bigg] \nn\\
&\times& \bigg[ \frac{-g}{4M} \frac{1}{ p^0 - q_n^0 - q_{n-1}^0 - q_{n-2}^0 - \frac{({\bs q}_n+{\bs q}_{n-1}+{\bs q}_{n-2})^2}{4M} + i\epsilon} \big(2{\bs q}_n+2{\bs q}_{n-1}+{\bs q}_{n-2} \big) \cdot {\bs A}(q_{n-2}) \bigg] \nn\\ [2pt]
&\times & \cdots 
\times \bigg[ \frac{-g}{4M} \frac{1}{ p^0 -  \sum_{j=1}^n q_j^0 - \frac{(\sum_{j=1}^n{\bs q}_j)^2}{4M} + i\epsilon } \Big(2\sum_{j=2}^{n}{\bs q}_j + {\bs q}_1 \Big) \cdot {\bs A}(q_1) 
\bigg] \nn\\ [4pt]
&=&   \bigg[ -g^2 \int \frac{\diff^3 q_{n}}{(2\pi)^3} \int \frac{\diff^3 q_{n-1}}{(2\pi)^3} \frac{1}{({\bs q}_n + {\bs q}_{n-1})^2 - i\epsilon} {\bs A}(t=+\infty, {\bs q}_{n}) \cdot  {\bs A}(t=+\infty, {\bs q}_{n-1}) \bigg] \nn\\
&\times& \bigg[\prod_{j=1}^{n-2} \int \frac{\diff^3 q_j}{(2\pi)^3} \bigg] 
\bigg[ g \frac{(2{\bs q}_n + 2{\bs q}_{n-1} + {\bs q}_{n-2}) \cdot {\bs A}(t=+\infty, {\bs q}_{n-2})}{({\bs q}_n+{\bs q}_{n-1}+{\bs q}_{n-2})^2 -i\epsilon} \bigg]  \nn\\
\label{eqn:A2}
&\times& \cdots \times \bigg[ g \frac{(2\sum_{j=2}^{n}{\bs q}_j + {\bs q}_1 ) \cdot {\bs A}(t=+\infty, {\bs q}_1)} {(\sum_{j=1}^n {\bs q}_j)^2-i\epsilon}  \bigg]
\ee

Plugging the Fourier transform (\ref{eqn:fourier}), doing a change of variables, ${\bs k}_m = \sum_{j=m}^n {\bs q}_j $ as done above and using (\ref{eqn:phase_trick}) we find the term in the third to last line of (\ref{eqn:A2}) leads to
\be
-g^2 \int \diff^3 R_{n-1} \frac{{\bs A}^2(t=+\infty, {\bs R}_{n-1})}{4\pi |{\bs R}_{n-1} - {\bs R}_{n-2}| } \,,
\ee
which cancels the second term in (\ref{eqn:second_order}) exactly.

Repeating the same procedure, one can show the $n$-th order (in $g$) diagram contributes 
\be
\frac{ (ig)^n}{n!} \phi^n({\bs R}=0)  \,,
\ee
where the position is set to the origin since the last delta function obtained by the above procedure is simply $\delta^3({\bs R}_1)$.

From (\ref{eqn:pure_gauge}), we find
\be
\phi({\bs R}=0) = - \int_0^\infty \diff {\bs R} \cdot {\bs A}(t=+\infty, {\bs R}) \,,
\ee
where we have used $\phi({\bs R}=\infty) = 0$. Using
\be
\nabla_{\bs R} \phi^n({\bs R}) = n\, \phi^{n-1}({\bs R})  \, \nabla_{\bs R}\phi({\bs R}) \,,
\ee
we can show
\be
\frac{1}{n!}\phi^n({\bs R}=0) &=& (-1)^n \int_0^\infty \diff {\bs R}_1 \cdot {\bs A}(t=+\infty, {\bs R}_1) \int_0^{{\bs R}_1} \diff {\bs R}_2 \cdot {\bs A}(t=+\infty, {\bs R}_2) \cdots \nn\\
&\times& \int_0^{{\bs R}_{n-1}} \diff {\bs R}_n \cdot {\bs A}(t=+\infty, {\bs R}_n) \,.
\ee
Summing over $n$ leads to
\be
\sum_n \frac{ (ig)^n}{n!} \phi^n(0) 
= \ml{P}\exp\Big( -ig\int_0^{\infty} \diff {\bs R} \cdot {\bs A}(t=+\infty,{\bs R}) \Big) \,.
\ee
The negative sign with respect to (\ref{eqn:wilson_line}) is due to the signature of the Lorentz metric.

In this derivation we chose ${\bs R}_0=0$ and thus the Wilson line starts at ${\bs R}_0=0$. For a nonvanishing ${\bs R}_0$, we need to restore the factor
\be
\exp\Big(i \sum_{j=1}^n{\bs q}_j \cdot {\bs R}_0 \Big) \,,
\ee
in (\ref{eqn:Mn}). Then the last delta function will become $\delta^3({\bs R}_1 - {\bs R}_0)$ and the Wilson line will start at ${\bs R}_0$.

The gauge link in the amplitude starts from the spatial position of the octet field and extend to spatial infinity. In the complex conjugate of the amplitude, the gauge link comes from the spatial infinity and stops at the spatial position of the other octet field. This leads to the Wilson lines at infinite time shown in Fig.~\ref{fig:wilson}.

\bibliographystyle{jhep}
\bibliography{main.bib}
\end{document}